\documentclass[journal,showpacs]{IEEEtran}
\usepackage{amsmath,amssymb,graphicx}
\usepackage{multirow}
\usepackage{epsf}
\usepackage{comment}

\usepackage{CJK}
\usepackage{amsmath}
\usepackage{indentfirst}%
\usepackage{graphicx}
\usepackage{float}
\usepackage{subcaption}
\usepackage{caption}
\begin{document}
%

\title{Secret key distillation over satellite-to-satellite free-space optics channel with a limited-sized aperture eavesdropper in the same plane of the legitimate receiver}
%
%

\author{Ziwen Pan and Ivan B. Djordjevic %
}

\markboth{Journal of \LaTeX\ Class Files,~Vol.~14, No.~18, August~2016}%
{Shell \MakeLowercase{\textit{et al.}}: Bare Demo of IEEEtran.cls for IEEE Journals}


\maketitle

\begin{abstract}
Conventionally, unconditional information security has been studied by quantum cryptography although the assumption of an omnipotent eavesdropper is too strict for some realistic implementations. In this paper, we study the realistic secret key distillation over a satellite-to-satellite free space optics channel where we assume a limited-sized aperture eavesdropper (Eve) in the same plane of the legitimate receiver (Bob) and determine the secret key rate lower bounds correspondingly. We first study the input power dependency without assumptions on Bob's detection scheme before optimizing the input power to determine lower bounds as functions of transmission distances, center frequency or Eve aperture radius. Then we calculate analytical expressions regarding the SKR lower bound and upper bound as transmission distance goes to infinity. We also incorporate specific discrete variable (DV) and continuous variable (CV) protocols for comparison. We demonstrate that significantly higher SKR lower bounds can be achieved compared to traditional unrestricted Eve scenario. 
\end{abstract}




\section{Introduction}

Quantum cryptography  theoretically promises unconditional informational security in physical layer.  The first quantum key distribution (QKD) scheme, BB84, was developed in 1984 by Charles Bennett and Gilles Brassard~\cite{bennett1984quantum} in which the security is based on no-cloning theorem and one-time pad encryption. This was the starting point for discrete variable QKD (DV-QKD) where single photons are transmitted, providing security while also 
requiring harsh conditions for realistic implementations, and various protocols have since been proposed~\cite{inoue2002differential,hwang2003quantum,pan2017quantum,PhysRevLett.108.130503,PhysRevLett.108.130502,PhysRevA.86.062319}. 

Nowadays with the uprising need for security in communication, people have been working to implement QKD in realistic application scenarios and thus continuous variable QKD (CV-QKD) has become an attractive field thanks to its advantages in experimental implementation, e.g., protocols based on coherent laser light and heterodyne detection~\cite{LPFPP18, DL15}. Recently, the longest-distance experimental demonstration of CV-QKD has been reported in~\cite{zhang2020long}, doubling the record in 2016~\cite{huang2016long}. However most research around the security of QKD assumes no restrictions on the eavesdropper (Eve) at all except the law of physics, which is not the case for realistic application. In our  papers~\cite{pan2019secret,8849223} we have presented the theoretical analysis of realistic secret key distillation by performing achievable rate calculation with certain restrictions to Eve's collecting ability, for example the aperture size of Eve's receiver if secret key distillation is implemented in a wireless communication channel. 

In recent years, with the development of 
satellite-based free-space communications~\cite{toyoshima2005trends}, 
the capacity and security of communication between satellites have become important. Thus interests have been rising surrounding free-space secret key distillation for satellites, especially since the work on satellite-to-ground quantum key distribution in 2017~\cite{liao2017satellite} and there have already been some interesting works analyzing such a unique application scenario~\cite{hayashi2020physical,vergoossen2019satellite}. 
With our results on the restricted eavesdropping model~\cite{pan2019secret} focusing on the limited collecting ability of Eve, which can easily apply on satellite based free space channels, as a continuation of~\cite{pan2020ICTON} we 
look into one of the typical scenarios where Eve's collecting ability is taken into consideration with the assumption of a limited-sized aperture for Eve.  
We assume that Eve's aperture is located in the same plane with Bob whereas the case when she can optimize her position after Bob is analyzed in~\cite{pan2020OSA}. 

In Section.~\ref{SamePlane} we introduce the problem setup of the limited-sized aperture of Eve in the same plane of Bob and calculate the channel parameters assuming that Gaussian beam is transmitted. Based on this, in Sec.~\ref{InputPowerDependency} we analyze the input power dependency of secure key rate (SKR) lower bounds with assumptions of different channel parameters.
Then in Sec.~\ref{LowerBoundswithOptimizedInputPower} we optimize input power based on the results from Sec.~\ref{InputPowerDependency}. We further investigate perfect and imperfect reconciliation schemes respectively in Secs.~\ref{PerfectReconciliationScheme} and \ref{ImperfectReconciliationScheme}.  We study the scenario where Eve's collecting ability is limited by the size of her aperture.  
We show that when Eve is in the same plane with Bob, as long as her aperture size is limited we will be able to achieve a non-zero secret key rate lower bound as transmission distance goes to infinity. We also present the comparison between Gaussian-modulated CV-QKD and decoy-state BB84 (DS-BB84) protocols under similar assumptions. 

\section{Limited-sized Aperture Scenario  Setup}\label{SamePlane}

\begin{figure}[htbp]
\centering{\includegraphics[height=13pc]{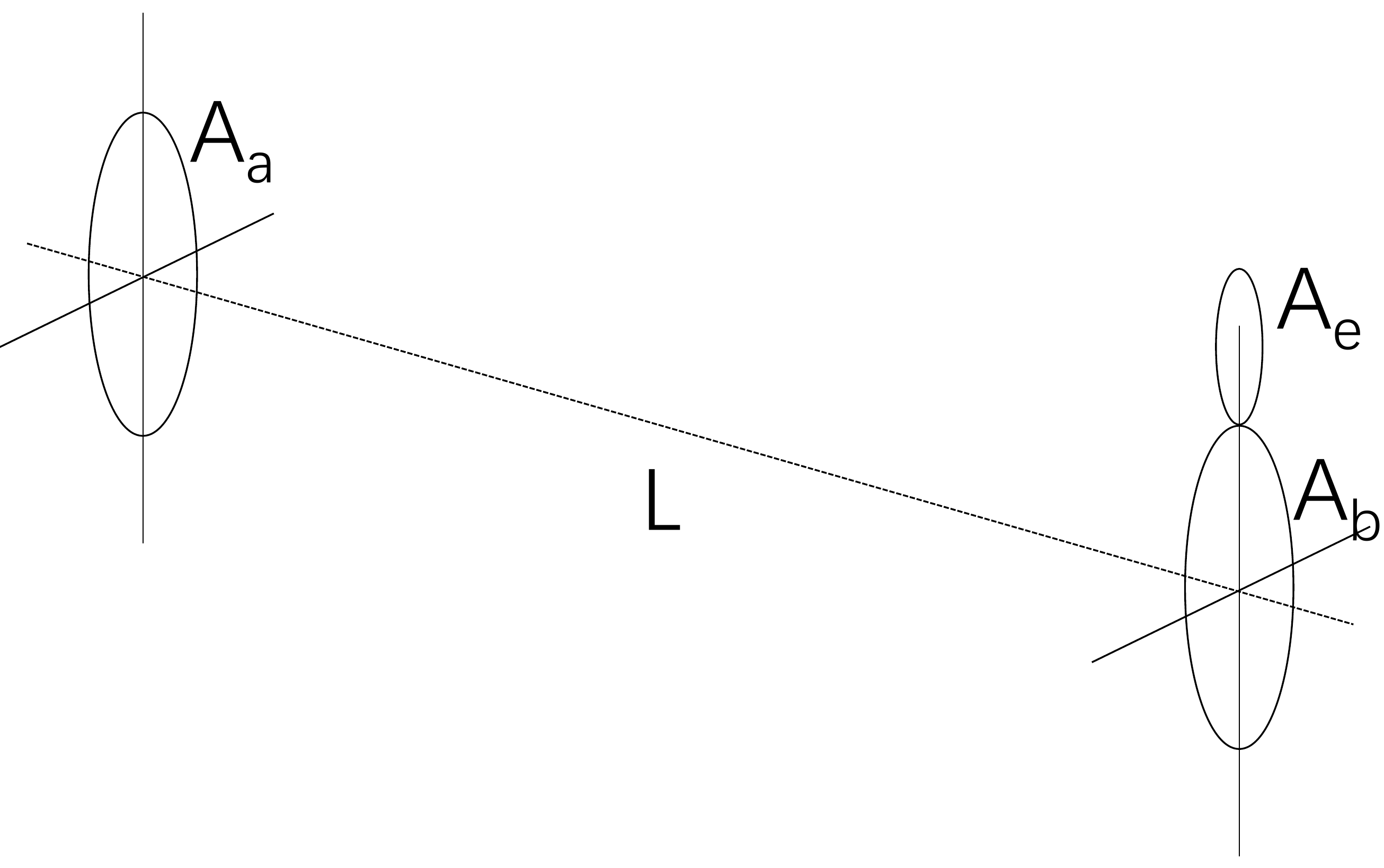}}
\caption{Geometric setup of the limited-sized aperture scenario. $A_a$ is the transmitting aperture (Alice) area and $A_b$ is the receiving aperture (Bob) area. $L$ is the  transmission distance between Alice and Bob. $A_{e}$ is the eavesdropper aperture area which is placed in the same plane as Bob's aperture. \label{EvePo1}}
\end{figure}

\noindent In this section we introduce the Limited Aperture Scenario, in which we assume that Eve has a limited-sized aperture 
$A_\text{Eve}$ ($A_e$) with radius $r_\text{Eve}$ ($r_e$), as in Fig.~\ref{EvePo1}. In this paper we only look at this straightforward case which actually gives us very interesting results when Eve's aperture is in the same plane as Bob's aperture with no overlapping. In this case, the optimal position is when Eve's aperture is tangential to Bob's aperture, 
as illustrated in Fig.~\ref{EvePo1}. Here we denote the area of transmitter aperture as $A_\text{Alice}$ ($A_a$) with radius $r_\text{Alice}$ ($r_a$) and the area of receiver aperture as $A_\text{Bob}$ ($A_b$) with radius $r_\text{Bob}$ ($r_b$).  
Since space optical communication usually uses near-infrared (NIR: 750nm to 1450nm) and short-infrared (SIR: 1400nm to 3000nm) wavelength ranges~\cite{kaushal2016optical}, we would mostly focus on 1550nm in this paper.

Although for conventional free space communication the receiver aperture diameter usually ranges from a few centimeters to even a few decimeters, to get an optimistic estimation 
on how large Eve's aperture can be, we take into consideration the latest technology level in manufacturing such a receiver aperture. Currently one of the largest optical observatories, the Giant Magellan Telescope, is designed around a 25-meter-diameter primary mirror with a collecting area of $386$m$^2$ consisting of seven 8.4-meter-diameter segments~\cite{johns2012giant}. However, we should also consider the limiting factors that come from the application scenario of satellite-to-satellite communication such as the difficulty of transporting such a large-sized receiver aperture into space 
and making it work. To get a fair reference on that, for instance, the Hubble space telescope, which works in the near ultraviolet, visible, and near infrared (approximately 0.1$\mu$m to 1 $\mu$m) spectra, has a 2.4-meter-diameter primary mirror at its center~\cite{montagnino1986test}. Its planned successor, the James Webb Space Telescope, which works at an extended wavelength range of 0.6$\mu$m to 27$\mu$m~\cite{lightsey2012james}, has a segmented primary mirror with its diameter at approximately 6.5m~\cite{clampin2008status}.  Some other devices designed to work mainly in infrared wavelength range use even smaller apertures. For example, the NASA infrared-wavelength astronomical space telescope "Wide-field Infrared Survey Explorer" uses a 40cm-diameter infrared telescope~\cite{wright2010wide}; the European space telescope "CHEOPS", which works at visible-near-infrared range of 330–1100nm, features an aperture of 30cm in effective diameter~\cite{hoyer2020expected}.   
For this work we will present the analysis using the typical aperture size for free space communications and study the potential rate using large-sized apertures. In reality, things would be more complex since Bob would surely try to increase his aperture as well to compete with Eve's increased aperture.

If we assume that the Gaussian beam is being transmitted in space, the  amplitude squared expression of the transmitted light beam can be expressed as:

\begin{equation}
    \|U(\rho,L)\|^2=I_0\left(\frac{W_0}{W(L)}\right)^2\exp\left({-\frac{2\rho^2}{W^2(L)}}\right).
\end{equation}
where $W_0$ is the waist radius of the Gaussian beam. Since generally a Gaussian beam with a larger waist radius requires a larger transmitter aperture, in the rest of this paper, without loss of generality, we assume $W_0=r_a$ to study how the transmitter aperture size would affect the achievable rates by affecting the beam diameter. Here $I_0$ is the intensity at the center of the beam at its waist, $L$ is the propagation distance with
\begin{align}
    z_0&=W_0^2\frac{\pi}{\lambda},\\
    W(L)&=W_0\sqrt{1+(L/z_0)^2},\\
    \rho^2&=x^2+y^2.
\end{align}
with $x, y$ being the transverse coordinates and $\lambda$ being the wavelength of the beam being transmitted, which is set to 1550nm in the rest of this paper unless specified otherwise. $z_0$ is the Rayleigh length for Gaussian beam propagating in free space.


Without loss of generality due to the symmetric power distribution on receiving plane, we assume that Eve puts her aperture right above Bob's aperture and perform the integration respectively to get each one's receiving power $P_{Bob}$ and $P_{Eve}$. We also assume Bob's aperture radius to be $r_b$ and Eve's one to be $r_e$.

Bob and Eve's received power can be calculated, respectively, as follows:

\begin{align}
    P_{Bob}&=I_0\int_{-r_b}^{r_b} \frac{\pi ^{3/2} W_0^3 e^{-\frac{2 \pi ^2 W_0^2 y^2}{A}} \text{erf}\left(\frac{\sqrt{2} \pi  W_0 \sqrt{r_b^2-y^2}}{\sqrt{A}}\right)}{\sqrt{2} \sqrt{A}} \, dy,\label{PBob}\\
    P_{Eve}&=\nonumber\\
    I_0&\int_{-r_e}^{r_e} -\frac{\pi ^{3/2} W_0^3 e^{-\frac{2 \pi ^2 W_0^2 x^2}{A}} E\left(\frac{\sqrt{2} \pi  W_0 C}{\sqrt{A}},\frac{\sqrt{2} \pi  W_0 B}{\sqrt{A}}\right)}{2 \sqrt{2} \sqrt{A}} \, dx.\label{PEve}
\end{align}

with 

\begin{align}
    A&=\pi ^2 W_0^4+\lambda ^2 L^2,\label{A}\\
    B&=\sqrt{r_e^2-x^2},\\
    C&=r_b+r_e,\\
    E(x,y)&=\text{erf}\left(x-y\right)-\text{erf}\left(x+y\right),
\end{align}

The total power in this Gaussian beam is:
\begin{equation}
    P_{total}=I_0\frac{\pi  W_0^2}{2},
\end{equation}

Now we can calculate Alice-to-Bob transmissivity ($\eta$) and Eve's fraction of collected power ($\kappa$)~\cite{pan2019secret} by: 
\begin{align}
    \eta&=\frac{P_{Bob}}{P_{total}},\label{eta}\\
    \kappa&=\frac{P_{Eve}}{(1-\eta)P_{total}}\label{kappa},
\end{align}
Also for noise frequency dependence we use the black body radiation equation with $n_e$ being the mean photon number per mode for the thermal noise:
\begin{equation}
n_e=\frac{1}{e^{\frac{hf}{kT}}-1}\label{blackradia}.
\end{equation}
where $h$ is the Planck constant, $k$ is the Boltzmann constant, $T=3\text{K}$ is the space temperature and $f$ is the center frequency in Hz that we use in transmission.

\section{Input Power Dependency}\label{InputPowerDependency}

\noindent In this section we investigate the input power dependency of SKR lower bounds based on Eqs.~(\ref{eta}), (\ref{kappa}) and (\ref{blackradia}). Recall from \cite{pan2019secret} that the lower bound for direct ($K_\rightarrow$) and reverse ($K_\leftarrow$) reconciliation respectively in a quantum thermal noise wiretap channel with reconciliation efficiency $\beta\in (0,1]$  are  given respectively as:
\begin{align}
    K_\rightarrow &\geq \beta g\left(n_e(1-\eta)+\eta\mu\right)-\sum_i g\left(\frac{\nu^{ER}_i-1}{2}\right)\nonumber\\
&-\beta g\left(n_e(1-\eta)\right)+g\left(n_e(1-\eta\kappa)\right)\label{LBDmu},\\
K_\leftarrow &\geq \beta g(\mu)-\sum_i g\left(\frac{\nu^{ER}_i-1}{2}\right)\nonumber\\
&-\beta g\left(\mu-\frac{\eta\mu(1+\mu)}{1+n_e-n_e\eta+\eta\mu}\right)+\sum_i g\left(\frac{\nu^{ER}_{y_i}-1}{2}\right)\label{LBRmu}.
\end{align}
where $g(x)$ is defined as:
\begin{align}
g(x)=(x+1)\log_2(x+1)-x\log_2(x).\label{thentropy}
\end{align}
Here $\mu$ is the average photon number that Alice transmits to Bob. If we make reasonable assumptions on $A_a$, $A_b$ and use $A_{e}$ as a parameter then we can plot direct (dashed curves) and reverse (solid curves) SKR lower bounds against input power $\mu$. Below in this section we use yellow curves to denote the case when Eve's aperture is larger than Bob's ($r_{e}>r_b$) while blue curves denote the case when Eve has a smaller aperture ($r_{e}<r_b$). The transmission distance is set to 10km. 




\begin{figure}[htbp]
\centering
\centering
\includegraphics[width=8.8cm]{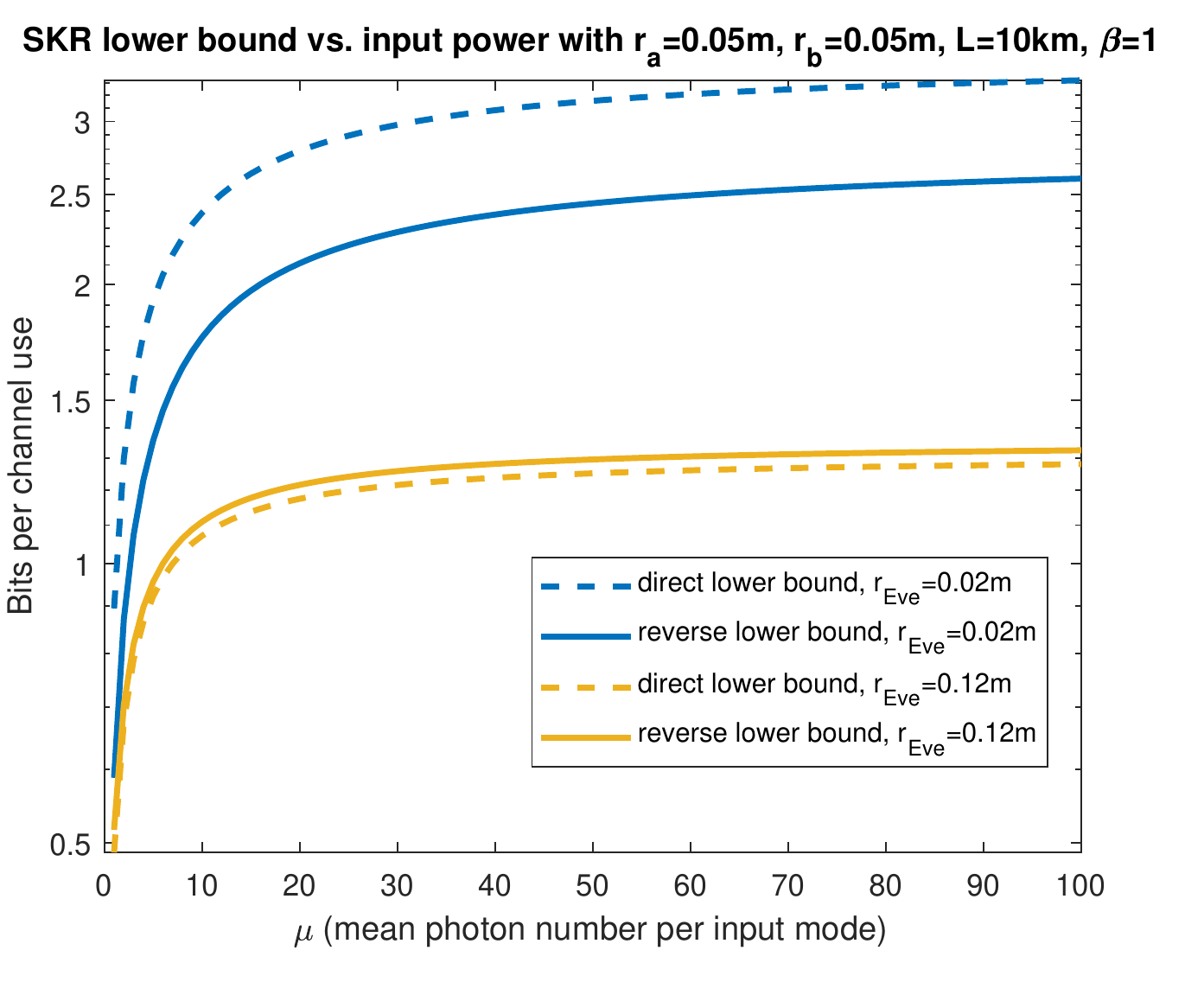}
\caption{SKR lower bound vs. input power. 
The transmission distance $L$ is 10km. Radius of Eve aperture  ($r_{e}$) are specified in the legend. Reconciliation efficiency $\beta$ is set to 1. Transmission center wavelength $\lambda$ is set to 1550nm. Transmitted Gaussian beam waist radius is set to $W_0=5$cm. Bob and Alice aperture radius are set to $r_b=r_a=5$cm. \label{LimitedApertureDR_mu_cropped}}
\end{figure}

In Fig.~\ref{LimitedApertureDR_mu_cropped} we present the lower bounds as functions of input power with different Eve aperture size when reconciliation is perfect ($\beta=1$). We can see that for both reconciliation schemes the SKR lower bounds increase with increasing input power, which means that the optimal input power is infinity when reconciliation is perfect ($\beta=1$). We can also see that when Eve's aperture is small ($r_e=2\text{cm}<r_b$), the direct reconciliation (blue dashed curve) can exceed the reverse reconciliation (blue solid curve). However when Eve's aperture increases ($r_e=12\text{cm}>r_b$), the direct reconciliation (yellow dashed curve) rate drops below reverse reconciliation (yellow solid curve). This is because increasing Eve's aperture only increases Alice-to-Eve transmissivity by increasing $\kappa$ without affecting Alice-to-Bob transmissivity $\eta$, which would decrease direct reconciliation rate more, similar to what we saw in~\cite{pan2019secret}.

\begin{figure}[htbp]
\centering
\centering
\includegraphics[width=8.8cm]{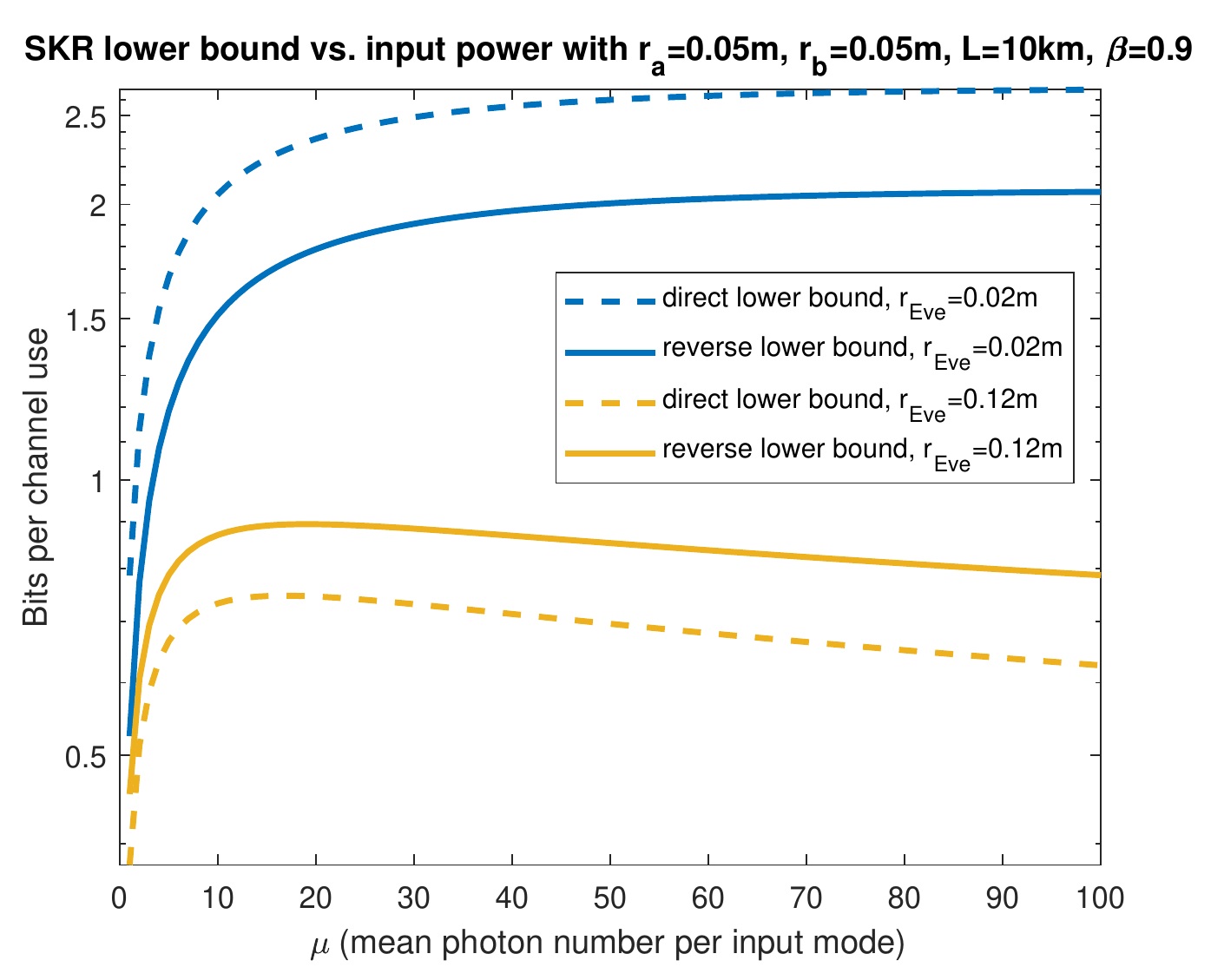}
\caption{SKR lower bound vs. input power. 
The transmission distance $L$ is 10km. Radius of Eve aperture  ($r_{e}$) are specified in the legend. Reconciliation efficiency $\beta$ is set to 0.9. Transmission center wavelength $\lambda$ is set to 1550nm. Transmitted Gaussian beam waist radius is set to $W_0=5$cm. Bob and Alice aperture radius are set to $r_b=r_a=5$cm. \label{LimitedApertureDR_mubeta9_cropped}}
\end{figure}

Next in Fig.~\ref{LimitedApertureDR_mubeta9_cropped} we set the reconciliation efficiency to 0.9. With $r_e=12$cm, compared to Fig.~\ref{LimitedApertureDR_mu_cropped} we can see a finite optimal input power for Alice when reconciliation is imperfect. However when Eve's aperture is small ($r_e=2\text{cm}<r_b$) we can obtain higher key rate exceeding this optimal input power. Similar to the perfect reconciliation case, here the direct reconciliation SKR lower bound (dashed curve) decreases below reverse reconciliation (solid curve) as Eve's aperture size increases.

\section{Lower Bounds with Optimized Input Power}\label{LowerBoundswithOptimizedInputPower}

\noindent In this section we optimize the input power and study the lower bounds as functions of transmission distance, center frequency and Eve's aperture size. Since in Sec.~\ref{InputPowerDependency} we have already seen that the optimized input power is different for different reconciliation efficiencies, we first study the case with perfect reconciliation in Sec.~\ref{PerfectReconciliationScheme} with input power taken to infinity. Then in Sec.~\ref{ImperfectReconciliationScheme} we optimize the input power accordingly with imperfect reconciliation.

\subsection{Perfect Reconciliation Scheme}\label{PerfectReconciliationScheme}

\begin{figure}[htbp]
\centering
\includegraphics[width=8.8cm]{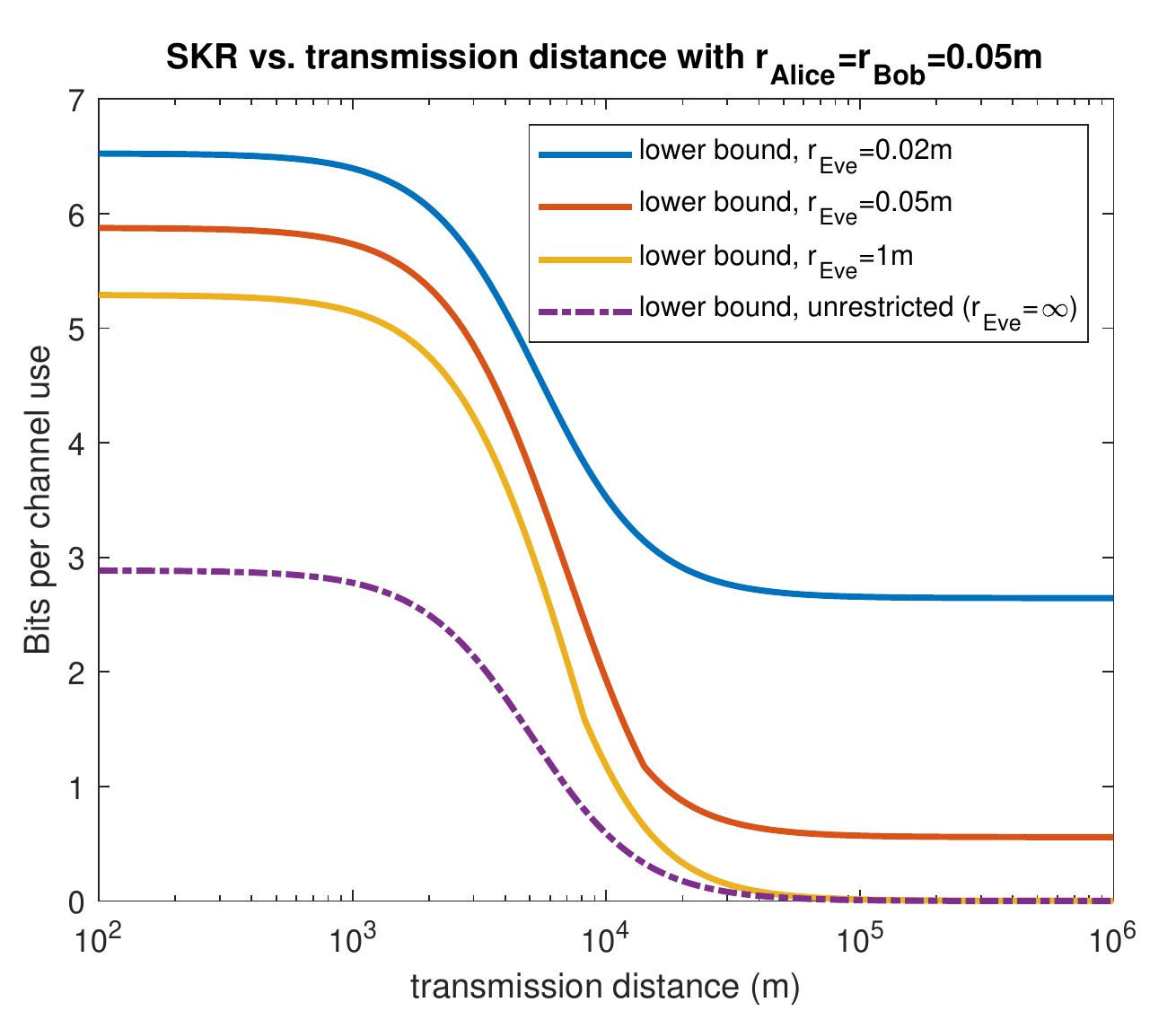}
\caption{SKR lower bound vs. transmission distance $L$ with optimized input power (infinity as reconciliation efficiency $\beta$ is set to 1). 
 Radius of Eve aperture  ($r_{e}$) are specified in the legend.  Transmission center wavelength $\lambda$ is set to 1550nm. Transmitted Gaussian beam waist radius is set to $W_0=5$cm. Bob and Alice aperture radius are set to $r_b=r_a=5$cm. 
Unrestricted Eve's case~\cite{pan2019secret} is also included for comparison. \label{ConstantRate}}
\end{figure}

From Fig.~\ref{LimitedApertureDR_mu_cropped} we conclude that the optimal input power is infinity when $\beta=1$. 
In  Fig.~\ref{ConstantRate} we plot the above SKR lower bounds with $\beta=1$ and $\mu\rightarrow\infty$ as the maximum of Eqs.~(\ref{LBDmu}) and (\ref{LBRmu}) against transmission distance assuming equal aperture sizes for Alice and Bob ($r_a=r_b=W_0=0.05\text{m}$). Here we assume the center wavelength of 1550nm. 

\begin{figure}[htbp]
\centering
\includegraphics[width=8.8cm]{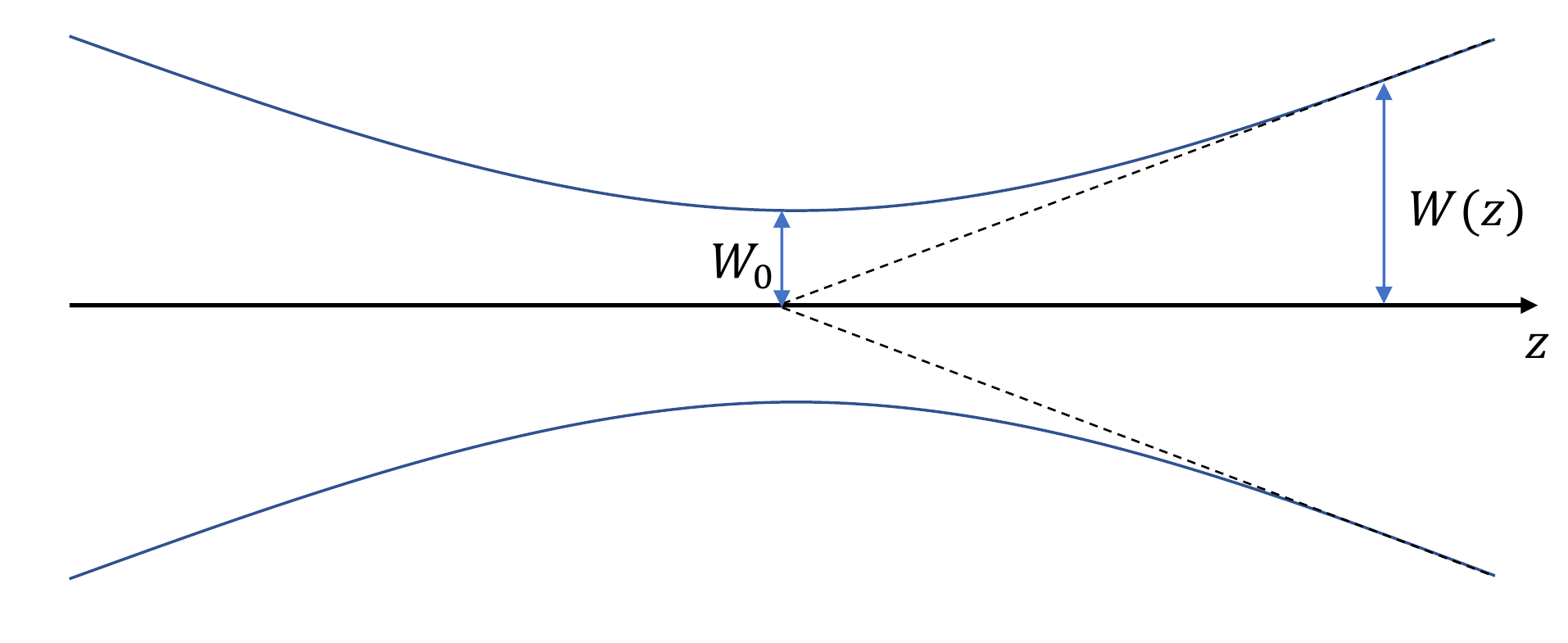}
\caption{Structure of Gaussian beam. Here $W_0$ is the beam waist radius, $W(z)$ is the Gaussian beam width as a function of the distance $z$ along the beam. \label{GaussianBeam}}
\end{figure}

Clearly, from Fig.~\ref{ConstantRate}  we can see that the SKR lower bounds converge to non-zero values either when transmission distance $L$ is sufficiently small or large. When $L$ is small, it's easy to see that converging trend as $L$ simply disappears from our Eqs.~(\ref{PBob}) and (\ref{PEve}) for $P_{Bob}$ and $P_{Eve}$ when we take $L\rightarrow0$, which thus gives us distance independent rates. This can be better understood through the structure of Gaussian beam as in Fig.~\ref{GaussianBeam}. We can see that the beam width decreases to the waist radius as approaching the focus of the beam ($z=0$). We can also see that as we approach the focus of the beam the beam width decreases more and more slowly to converge to the waist radius $W_0$. This leads to the non-zero rate when $L$ is small. 

To understand the constant rate when $L$ is large, we notice that the channel condition here actually approximates a pure loss channel as $n_e\approx 0$ in space ($T=3\text{K}$) at this wavelength. 
Recall from \cite{pan2019secret} that the lower bound for direct ($K_\rightarrow$) and reverse ($K_\leftarrow$) reconciliation respectively in a pure loss wiretap channel without transmitting power  and reconciliation efficiency constraints ($\beta=1$, $\mu\rightarrow\infty$) are given respectively as: 
\begin{align}
\centering
\lim_{\mu\rightarrow \infty, n_e\rightarrow 0}K_\rightarrow(\beta=1) &\ge \log_2\frac{\eta}{\kappa(1-\eta)},\label{LBD}\\
\lim_{\mu\rightarrow \infty, n_e\rightarrow 0}K_\leftarrow(\beta=1) &\ge \log_2\frac{1}{\kappa(1-\eta)}\nonumber\\
&-\left(g\left(\frac{1-\eta}{\eta}\right)-g\left(\frac{(1-\eta)\kappa}{\eta}\right)\right).\label{LBR}
\end{align}
In Fig.~\ref{ConstantRate} we can see that the SKR lower bounds first decrease with increasing distance.
When Eve has an infinite-sized aperture (unrestricted case in Fig.~\ref{ConstantRate}), the SKR would drop to zero as distance goes to infinity since increased distance only decreases Bob's collecting ability with no impact on Eve. 
However when Eve's aperture size is limited, the SKR stops decreasing and stays constant at a non-zero value as distance goes to infinity. This is because when distance is large, we will have the collecting ability of Bob and Eve being proportional to their aperture size:
\begin{align}\lim_{L\rightarrow\infty}\frac{P_{Eve}}{P_{Bob}}&=\lim_{L\rightarrow\infty}\frac{\eta_{AE}}{\eta_{AB}}=\frac{A_{e}}{A_b},\label{AeAb}\\
\eta_{AE}&=(1-\eta_{AB})\kappa\label{etakappa},
\end{align} 
where $\eta_{AB}$ and $\eta_{AE}$ refer to Alice-to-Bob and Alice-to-Eve transmissivity, respectively. 

Now if we define the ratio of Eve's and Bob's aperture as $m $:
\begin{equation}
\lim_{L\rightarrow\infty}\frac{\eta_{AE}}{\eta_{AB}}=\frac{A_{e}}{A_b}=\left(\frac{r_{e}}{r_b}\right)^2=m.\label{AeAbm}
\end{equation}

Relating to Eqs.~(\ref{LBD}) and (\ref{LBR}), this would return a non-zero SKR lower bound as $\eta_{AB}$ goes to zero (transmission distance $L$ goes to infinity): 
\begin{align}
\lim_{\mu\rightarrow \infty, L\rightarrow\infty}K_\rightarrow&\geq-\log_2(m),\label{LBD_SP_Linf}\\
\lim_{\mu\rightarrow \infty, L\rightarrow\infty}K_\leftarrow&\geq-\log_2\left(\left(\frac{m}{1+m}\right)^{1+m}e\right),\label{LBR_SP_Linf}\\
\lim_{\mu\rightarrow \infty, L\rightarrow\infty}K&\geq\max\left(K_\rightarrow,K_\leftarrow\right).\label{LBO_SP_Linf}
\end{align}

It's easy to see that when $A_e<A_b$ ($0<m<1$), meaning that Bob's collecting ability is higher than Eve's, we can have positive SKR lower bound for both direct ($K_\rightarrow$) and reverse ($K_\leftarrow$) reconciliation which would go to infinity when $A_e<<A_b$ ($m\rightarrow0$). However when $A_e>A_b$ ($m>1$), meaning that Eve's collecting ability is higher than Bob's, the lower bound on the direct reconciliation SKR would go to zero (SKR cannot take negative values) whereas the reverse reconciliation lower bound remains positive. When $A_e>>A_b$ ($m\rightarrow\infty$) the reverse reconciliation lower bound goes to zero as $\lim_{m\rightarrow\infty}\left(\frac{m}{1+m}\right)^{1+m}=\frac{1}{e}$.

In Fig.~\ref{ConstantRate} we can also see that when Eve's aperture is much larger than Bob's ($r_e=1\text{m}$), the SKR lower bound approximates the unrestricted Eve's case when distance is sufficiently large. However, when Eve's aperture is smaller than Bob's ($r_e=0.02\text{m}$), or even when they are equal ($r_e=0.05\text{m}$), the SKR tends to a non-zero value almost independent on transmission distance. 

Below in Fig.~\ref{FlatRateDR} we include lower bounds for both direct and reverse reconciliation with different Eve's aperture sizes. Here we use dashed curves to denote direct reconciliation and solid curves to denote reverse reconciliation. The unrestricted case lower bound, which approximates capacity since $n_e\approx 0$ in this case, is given with diamond markers for comparison. It's clear from Fig.~\ref{FlatRateDR} that the reverse reconciliation lower bounds always become distance independent when transmission distance is sufficiently large regardless of the aperture size of Eve. When $r_e>r_b$ ($r_e=0.07\text{m}$), direct reconciliation lower bound shows a limit in its transmission distance as its rate drops to zero at a finite transmission distance. When $r_e=r_b$ ($r_e=0.05\text{m}$), direct reconciliation lower bound becomes distance dependent and start to decrease with increasing distance in a similar manner to the unrestricted case. When $r_e=0.02\text{m}$, we can see that direct reconciliation lower bound is also distance independent after a certain transmission distance, and is even higher than reverse reconciliation. To get the exact threshold for $m $ when direct reconciliation lower bound exceeds that of reverse reconciliation, we compare Eq.~(\ref{LBD_SP_Linf}) and Eq.~(\ref{LBR_SP_Linf}) and conclude that for $m_{th}$ satisfying $(m_{th}+1)\log(m_{th}+1)-m_{th}\log(m_{th})=1$ we have the direct reconciliation lower bound higher than that of reverse reconciliation when $m<m_{th}$. We can numerically solve $m_{th}\approx0.54$. Thus from Eqs.~(\ref{LBD_SP_Linf}), (\ref{LBR_SP_Linf}) and (\ref{LBO_SP_Linf}) we conclude:
\begin{align}
\lim_{\mu\rightarrow \infty, L\rightarrow\infty}K&\geq\left\{
             \begin{array}{lr}
             -\log_2(m), & (m\leq m_{th}) \\
             -\log_2\left(\left(\frac{m}{1+m}\right)^{1+m}e\right).&
              (m>m_{th})\end{array}
\right.\label{LBOS_SP_Linf}\\
(m_{th}+1)&\log(m_{th}+1)-m_{th}\log(m_{th})=1
\end{align}

\begin{figure}[htbp] 
\centering{\includegraphics[height=17pc]{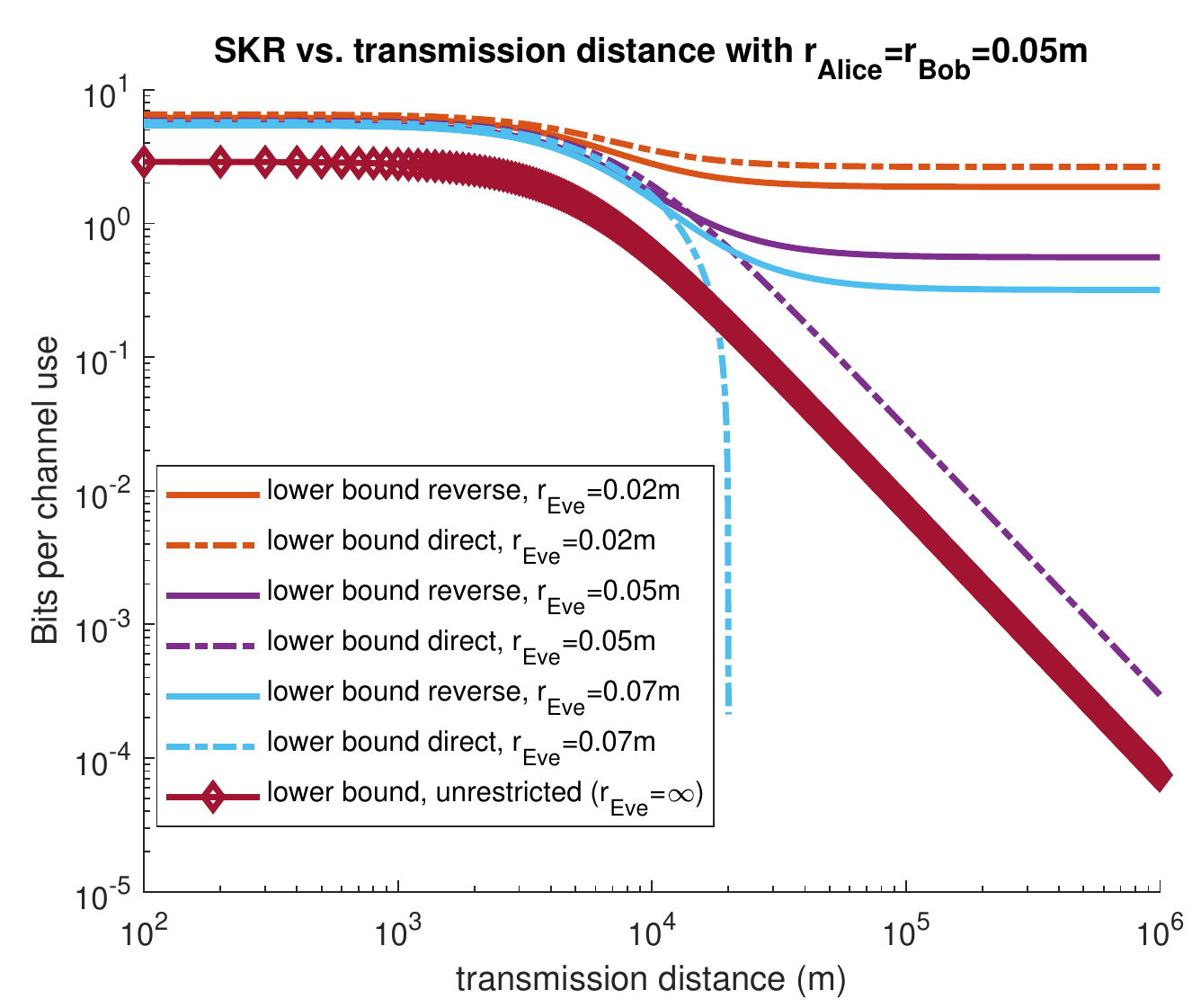}}
\caption{Direct and reverse reconciliation lower bounds  versus transmission distance $L$ with optimized input power (infinity as reconciliation efficiency $\beta$ is set to 1). 
 Transmission center wavelength $\lambda$ is set to 1550nm. Transmitted Gaussian beam waist radius is set to $W_0=5$cm. Bob and Alice aperture radius are set to $r_b=r_a=5$cm.  Radius of Eve aperture  ($r_{e}$) are varied as given. Unrestricted Eve ($r_e=\infty$) is also included for comparison.
 \label{FlatRateDR}}
\end{figure}

From~\cite{pan2019secret} we recall that the upper bound for pure loss channel ($n_e=0$) is of the form:

\begin{equation}
    \lim_{\mu\rightarrow \infty}\max\{K_\rightarrow,K_\leftarrow\}\le \log_2\left(\frac{\eta+\kappa(1-\eta)}{\kappa(1-\eta)}\right).\label{UpBREEPL}
\end{equation}
With Eqs.~(\ref{AeAb}), (\ref{etakappa}) and (\ref{AeAbm}) we can rewrite the above Eq.~(\ref{UpBREEPL}) as:
\begin{align}
   \lim_{\mu\rightarrow \infty, L\rightarrow\infty} \max\{K_\rightarrow,K_\leftarrow\}&\le \log_2\left(\frac{\eta+\kappa(1-\eta)}{\kappa(1-\eta)}\right)\\
   &\leq\log_2\left(\frac{1+m}{m}\right).\label{UpBREEPLm}
\end{align}

\begin{figure}[htbp] 
\centering{\includegraphics[height=17pc]{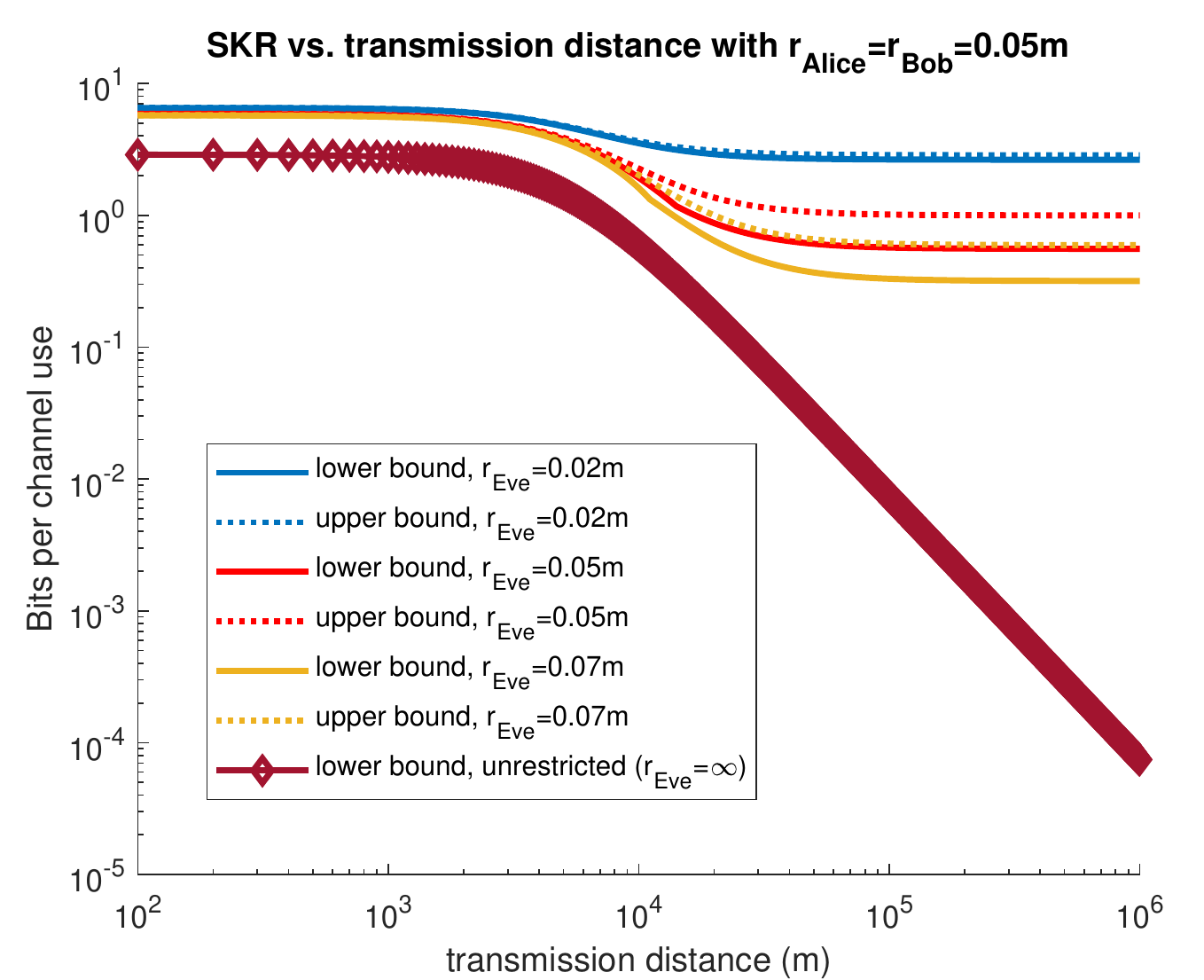}}
\caption{Lower bounds versus upper bounds against transmission distance $L$ with optimized input power (infinity as reconciliation efficiency $\beta$ is set to 1). 
 Transmission center wavelength $\lambda$ is set to 1550nm. Transmitted Gaussian beam waist radius is set to $W_0=5$cm. Bob and Alice aperture radius are set to $r_b=r_a=5$cm.  Radius of Eve aperture  ($r_{e}$) are varied as provided in the legend. Unrestricted Eve case ($r_e=\infty$) is also included for comparison.
 \label{FlatRateLvsU}}
\end{figure}

In Fig.~\ref{FlatRateLvsU}, we include the above upper bounds with dotted curves since $n_e\approx0$ in this plot. We also take the maximum of direct and reverse reconciliation lower bound as in Eq.~(\ref{LBOS_SP_Linf}) to plot as the lower bound in Fig.~\ref{FlatRateLvsU} in solid curves. Again, the unrestricted case lower bound, which approximates capacity since $n_e\approx 0$ in this case, is given with diamond markers for comparison. We can see that when $r_e<r_b$ ($r_e=0.02\text{m}$) the upper bound and lower bound are very close since direct reconciliation lower bound is higher than reverse when $m<m_{th}\approx0.54$ and  the upper bound $\log_2\left(\frac{1+m}{m}\right)$ starts to converge to the direct reconciliation lower bound $-\log_2(m)$ when $m<1 \text{ \& } m\rightarrow0$. Compared with  curves in which $r_e\geq r_b$ ($r_e=0.05\text{m}$ and $r_e=0.07\text{m}$) we can see that this gap between the lower bound and upper bound increases with increasing $m $, which is equal to relatively increasing Eve's collecting ability in comparison to Bob's.

Below in Fig.~\ref{SamePlanevsFrequency10kmDR} we plot the direct and reverse reconciliation lower bounds versus center frequency whereas the transmission distance is fixed as 10km. We can see from Fig.~\ref{SamePlanevsFrequency10kmDR} that for any given center frequency reverse reconciliation always has a positive lower bound whereas direct reconciliation lower bound drops to zero when frequency decreases below certain value with $r_e>r_b$ ($r_e=0.7\text{m}$ in Fig.~\ref{SamePlanevsFrequency10kmDR}). When $r_e=r_b$ ($r_e=0.05\text{m}$ in Fig.~\ref{SamePlanevsFrequency10kmDR}), direct reconciliation lower bound increases with increasing center frequency in a similar manner to the unrestricted case. And when $r_e<r_b$  ($r_e=0.01\text{m}$ in Fig.~\ref{SamePlanevsFrequency10kmDR}) direct reconciliation also has a positive lower bound regardless of the center frequency. Generally the achievable rates increase with increasing center frequency as increased center frequency decreases the beam divergence and thus decrease the power that can be collected by Eve.

\begin{figure}[htbp] 
\centering{\includegraphics[height=17pc]{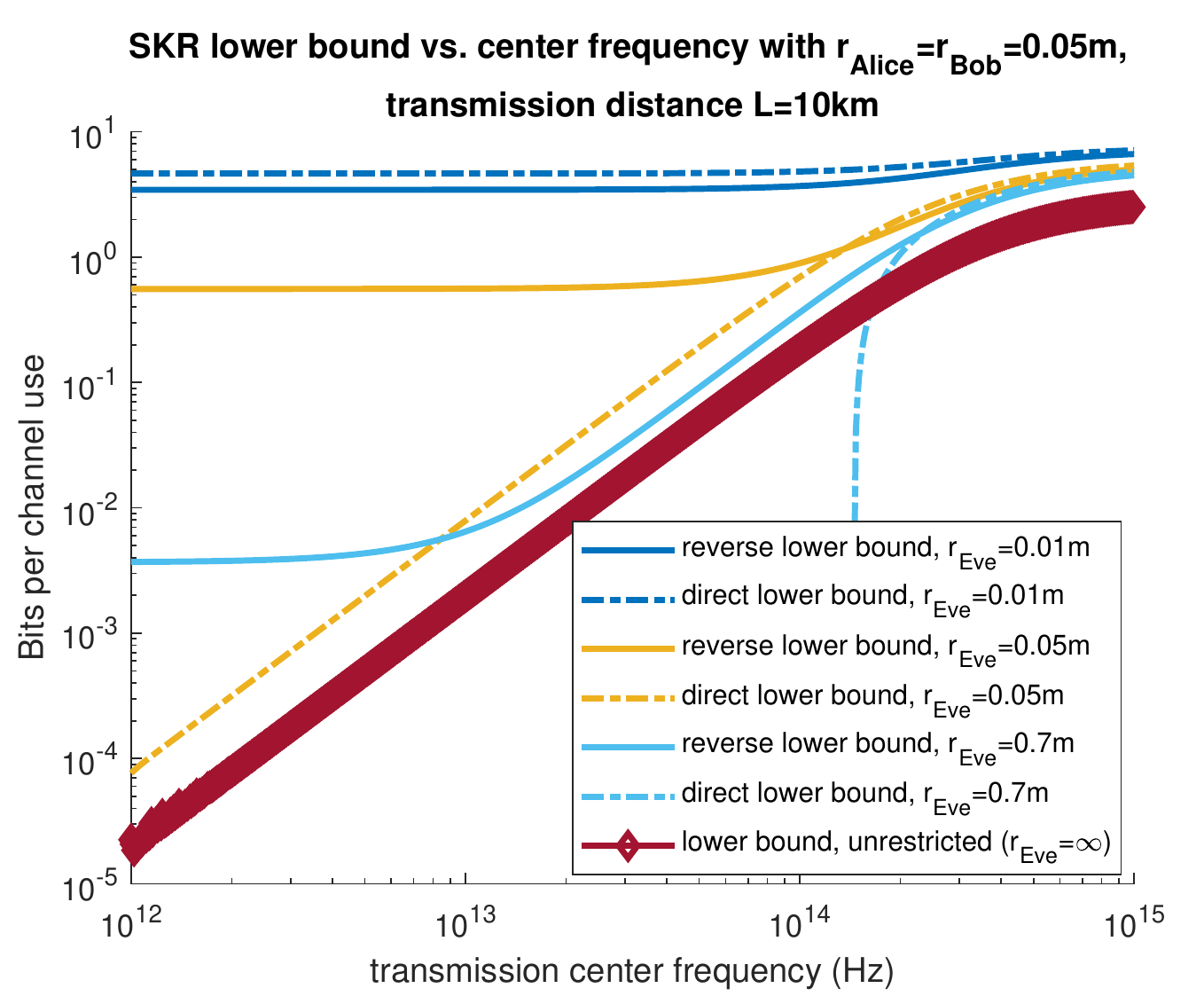}}
\caption{Direct and reverse reconciliation lower bounds as functions of transmission center frequency $f$ with optimized input power (infinity as reconciliation efficiency $\beta$ is set to 1). 
 Transmission distance $L$ is fixed as 10km. Transmitted Gaussian beam waist radius is set to $W_0=5$cm. Bob and Alice aperture radius are set to $r_b=r_a=5$cm.  Radius of Eve aperture  ($r_{e}$) are varied as given. Unrestricted Eve ($r_e=\infty$) is also included for comparison.
 \label{SamePlanevsFrequency10kmDR}}
\end{figure}

\begin{figure}[htbp] 
\centering{\includegraphics[height=17pc]{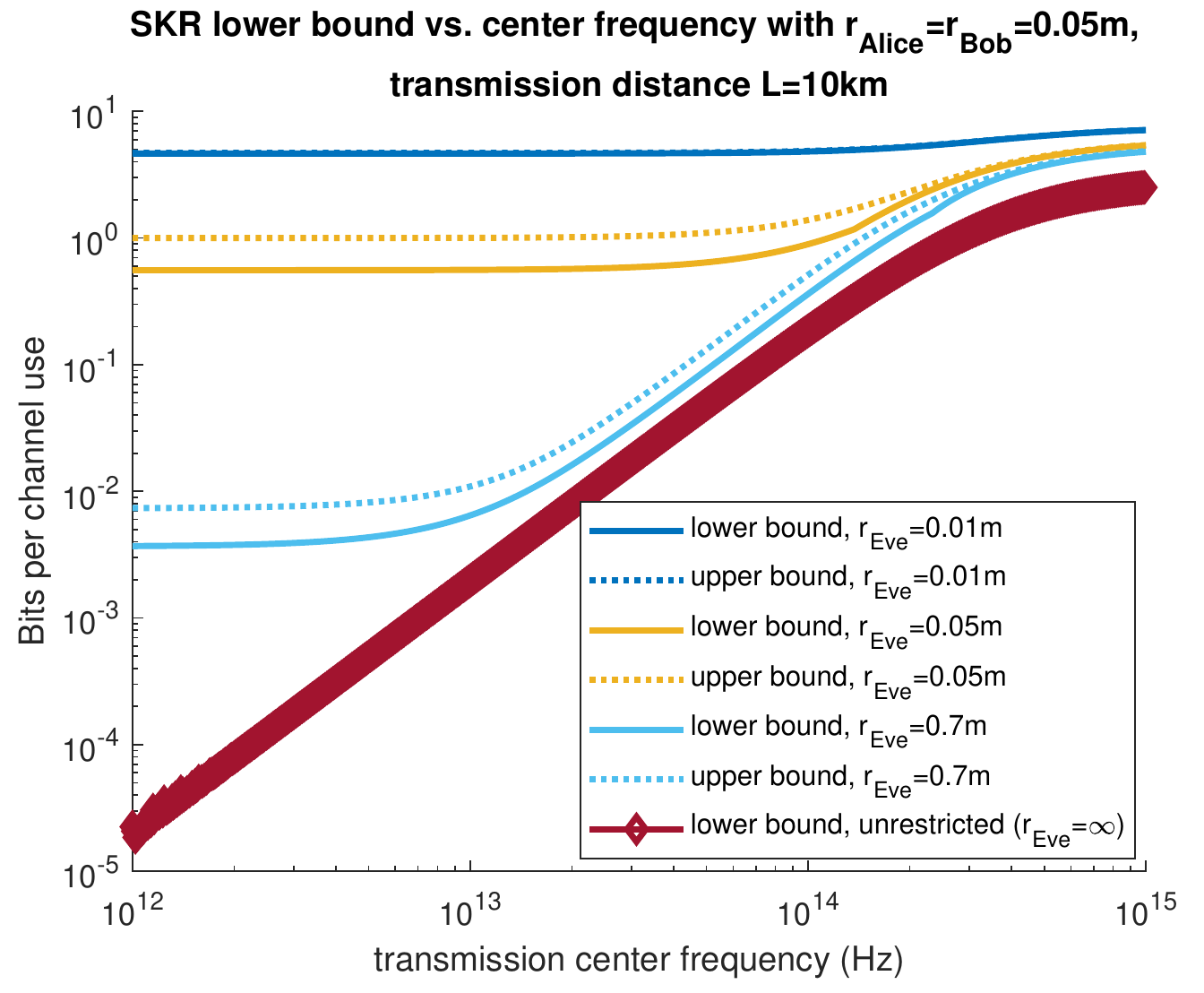}}
\caption{Lower bounds and upper bounds as functions of transmission center frequency $f$ with optimized input power (infinity as reconciliation efficiency $\beta$ is set to 1). 
 Transmission distance $L$ is fixed as 10km. Transmitted Gaussian beam waist radius is set to $W_0=5$cm. Bob and Alice aperture radius are set to $r_b=r_a=5$cm.  Radius of Eve aperture  ($r_{e}$) are varied as given. Unrestricted Eve ($r_e=\infty$) is also included for comparison.
 \label{SamePlanevsFrequency10km}}
\end{figure}

Next in Fig.~\ref{SamePlanevsFrequency10km} we include the upper bound (Eq.~(\ref{UpBREEPL})) and the lower bound (maximum of Eq.~(\ref{LBD}) and Eq.~(\ref{LBR})) versus the transmission center frequency whereas transmission distance is fixed as 10km. We can see that regardless of Eve's aperture size, both the lower bound and upper bound increase with increasing center frequency and start to converge to each other once the center frequency is sufficiently high and becomes frequency independent thereafter. Furthermore the gap between lower bound and upper bound decreases when Eve's aperture size decreases just like what we observed from Fig.~\ref{FlatRateLvsU}.

To have a better understanding of this "frequency independent" SKR lower bound that showed up in   Fig.~\ref{SamePlanevsFrequency10km} and Fig.~\ref{SamePlanevsFrequency10kmDR} when $f$ is either too low or too high, we notice that when frequency is sufficiently high, if we take center frequency $f\rightarrow\infty$ (equivalently $\lambda\rightarrow0$) in Eqs.~(\ref{PBob}) and (\ref{PEve}) then we would get rid of both wavelength $\lambda$ and transmission distance $L$ from the expression of SKR lower bound. This would explain why the SKR lower bound becomes frequency independent when frequency is sufficiently high since setting $f\rightarrow\infty$ is equal as setting $L=0$ with $\lambda$ and $L$ mathematically bound together through Eq.~(\ref{A}).

Similarly, when frequency $f$ is low in Fig.~\ref{SamePlanevsFrequency10km} the SKR lower bounds and upper bounds also become frequency independent. And since $f\rightarrow0$ ($\lambda\rightarrow\infty$) is equal as $L\rightarrow\infty$ we have the frequency independent lower and upper bound consistent with Eq.~(\ref{LBOS_SP_Linf}) and Eq.~(\ref{UpBREEPLm}), rewritten here as:
\begin{align}
\lim_{\mu\rightarrow \infty, \lambda\rightarrow\infty}K&\geq\left\{
             \begin{array}{lr}
             -\log_2(m), & (m\leq m_{th}) \\
             -\log_2\left(\left(\frac{m}{1+m}\right)^{1+m}e\right).&
              (m>m_{th})\end{array}
\right.\label{LBOS_SP_Linf_f}\\
\lim_{\mu\rightarrow \infty, \lambda\rightarrow\infty} K&\le \log_2\left(\frac{1+m}{m}\right).\label{UpBREEPLm_f}
\end{align}

\begin{figure}[htbp] 
\centering{\includegraphics[height=17pc]{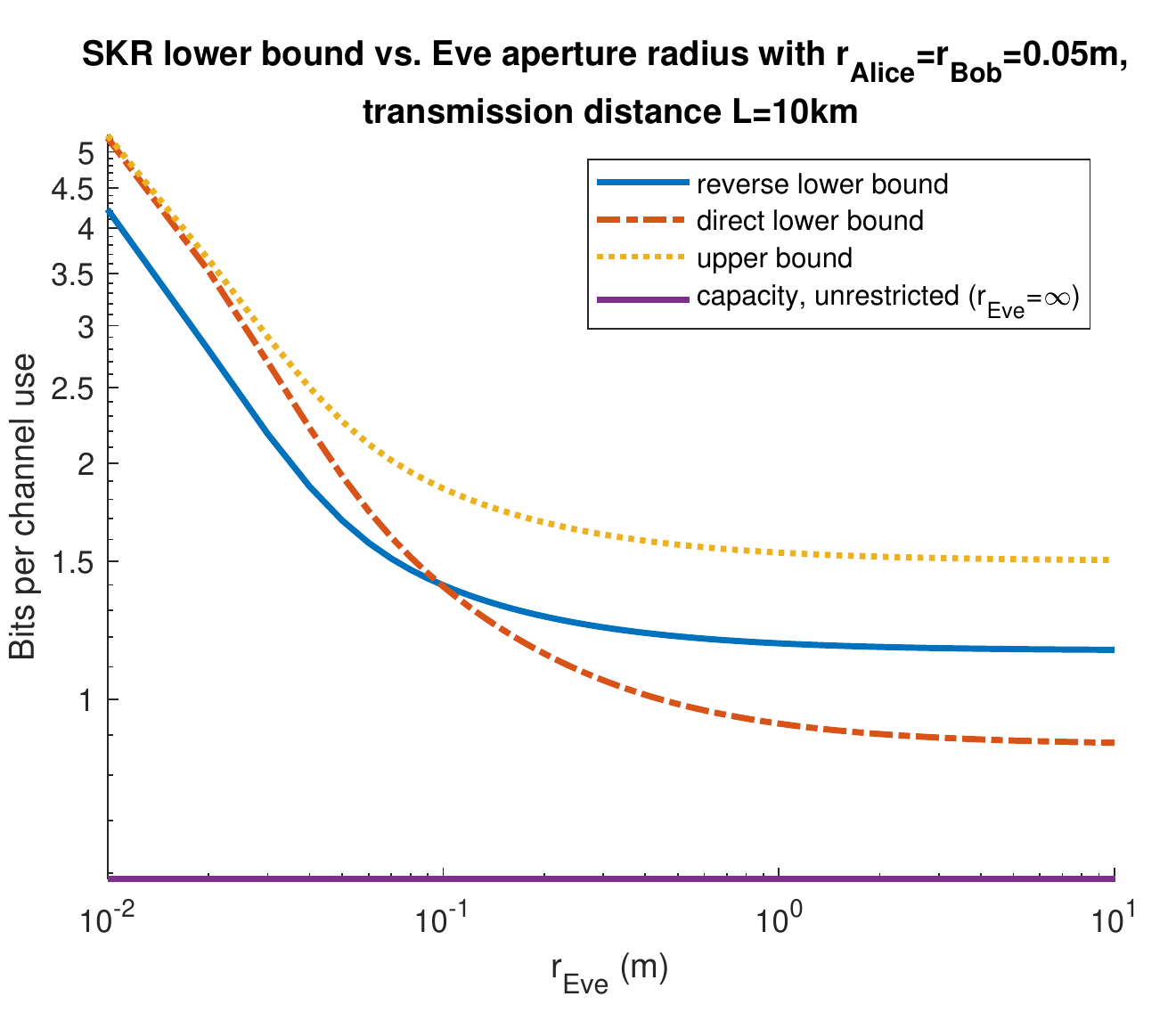}}
\caption{Lower bounds and upper bounds as functions of functions of Eve aperture radius $r_e$ with optimized input power (infinity as reconciliation efficiency $\beta$ is set to 1). 
 Transmission distance $L$ is fixed as 10km. Transmission center wavelength is set as $\lambda=1550$nm. Transmitted Gaussian beam waist radius is set to $W_0=5$cm. Bob and Alice aperture radius are set to $r_b=r_a=5$cm.   Unrestricted Eve ($r_e=\infty$) is also included for comparison.
\label{geo3222vre_cropped}}
\end{figure}

In Fig.~\ref{geo3222vre_cropped} we present the direct and reverse reconciliation lower bounds with upper bound and unrestricted case capacity included as functions of Eve's aperture radius. Here transmission distance is set to 10km, center wavelength $\lambda=$1550nm. We can see that when Eve's aperture size is small the direct reconciliation rate is higher and approaches the upper bound. However, when Eve's aperture size increases the direct reconciliation rate decreases until is lower than reverse reconciliation rate. Both reconciliation rates exceed the capacity of unrestricted Eve's case. We can also see that when Eve's aperture size further increases, the SKR lower and upper bounds start to converge to a constant rate as the further increased Eve's aperture would only be able to collect more light in the area farther away from the beam center where the beam intensity is almost negligible so that increasing Eve's aperture size does not increase Eve's collected light as much as when $r_e$ is small.

\begin{figure}[htbp]
\centering
\includegraphics[width=8.8cm]{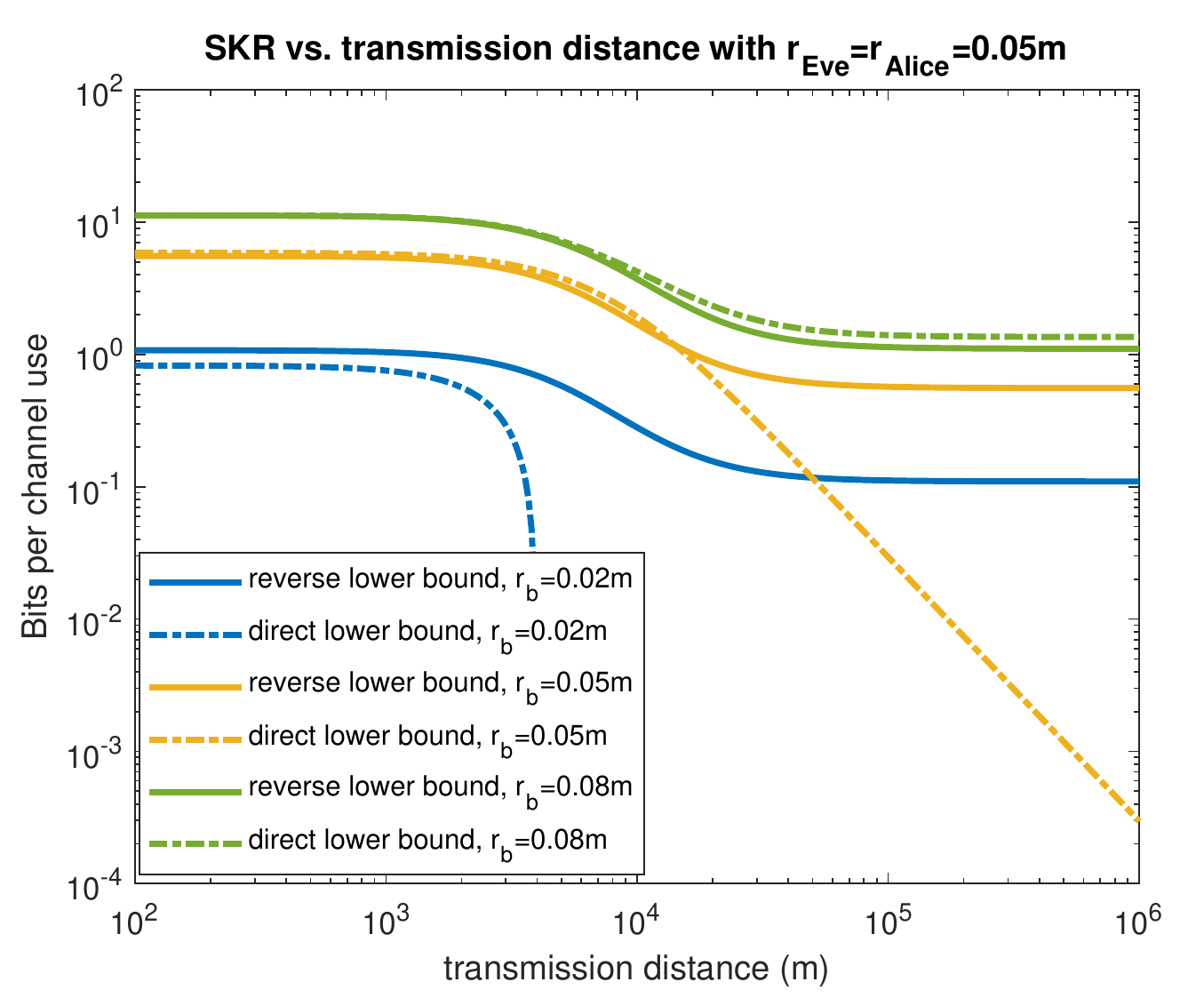}
\caption{Direct and reverse reconciliation SKR lower bounds vs. transmission distance $L$ with optimized input power (infinity as reconciliation efficiency $\beta$ is set to 1). 
 Radii of transmitted Gaussian beam waist are specified in the legend.  Transmission center wavelength $\lambda$ is set to 1550nm.  Eve and Alice aperture radius are set to $r_e=r_a=5$cm. 
 \label{ConstantRatewithrb}}
\end{figure}

In the above plots we mainly introduced how Eve's aperture size can affect the secure key rate lower and upper bounds. In reality, to compete with Eve, Alice and Bob would certainly try to increase their own aperture sizes in order to achieve higher secure key rates. So here we look at how Alice's aperture size, which mainly affects the transmitted Gaussian beam width in this paper, and Bob's aperture size, which affects the legitimate communication party's receiving ability, would affect the secure key rate lower and upper bounds. The case where only Bob's aperture size is increased is pretty obvious to see since this would increase Bob's collecting ability while diminishing Eve's, so the SKR achievable rate would certainly increase for any given transmission distance, as in Fig.~\ref{ConstantRatewithrb}.
\begin{figure}[htbp]
\centering
\includegraphics[width=8.8cm]{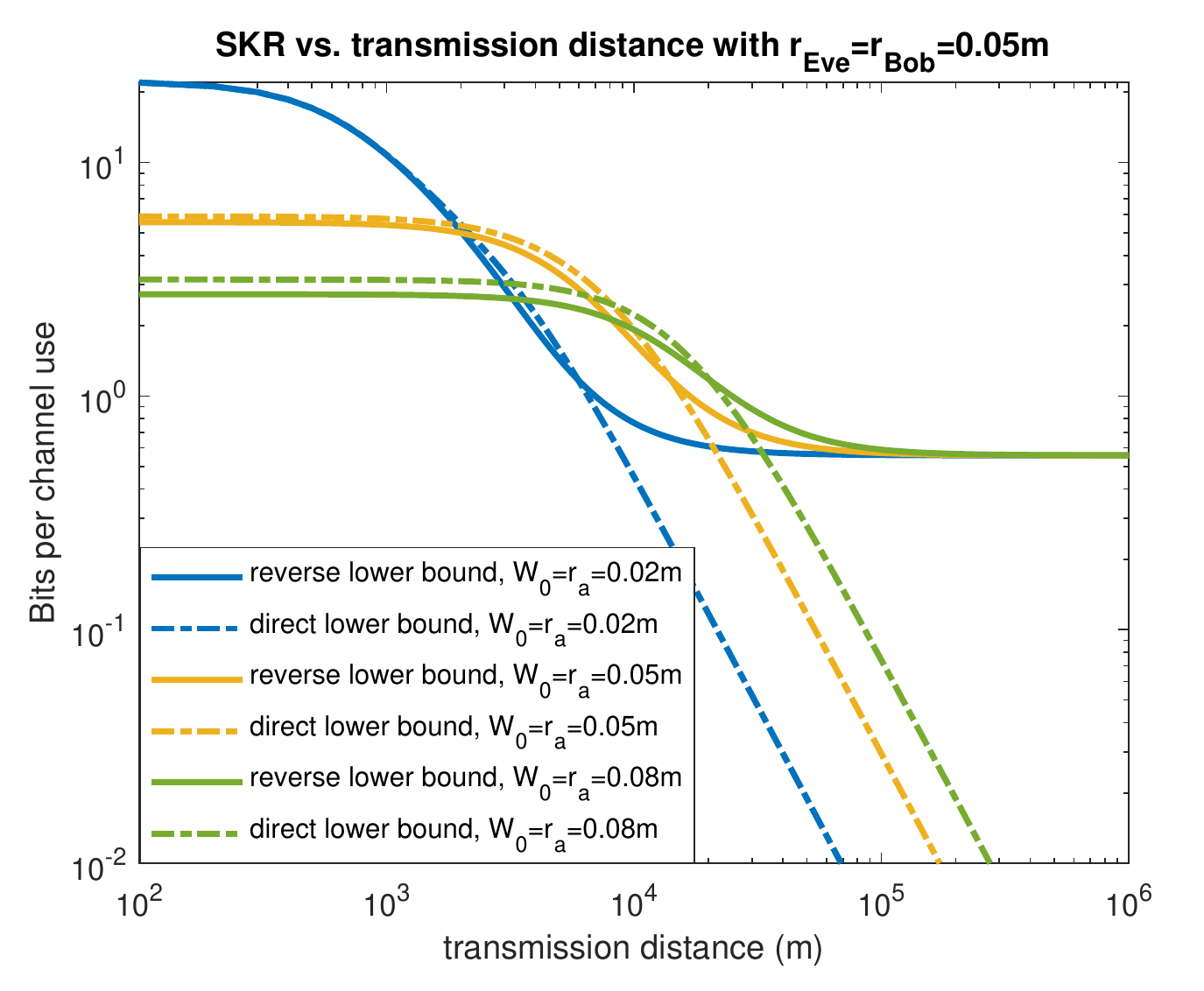}
\caption{Direct and reverse reconciliation SKR lower bounds vs. transmission distance $L$ with optimized input power (infinity as reconciliation efficiency $\beta$ is set to 1). 
 Radii of transmitted Gaussian beam waist are specified in the legend.  Transmission center wavelength $\lambda$ is set to 1550nm.  Eve and Bob aperture radius are set to $r_e=r_b=5$cm. 
 \label{ConstantRatewithra}}
\end{figure}

In Fig.~\ref{ConstantRatewithra} we look at the case where Bob's and Eve's aperture sizes are fixed and see how the transmitted Gaussian beam waist radius would affect both direct and reverse reconciliation lower bounds. We can see  that  when transmission distance is short direct reconciliation achievable rates are higher than reverse reconciliation achievable rates and when transmission distance is large the direct reconciliation achievable rates decrease with increasing transmission distance in a similar manner as we saw in Fig.~\ref{FlatRateDR} when $r_b=r_e$. On the other hand, the reverse reconciliation achievable rates converge to the same constant rate as here $r_e$ and $r_b$ remain the same. We can also see that when transmission distance is short, although a Gaussian beam with a smaller waist radius can achieve higher SKR lower bounds, especially when $r_b<W_0$ which makes it easier for Bob to collect most of the photons in the transmitted beam, as $L$ increases the achievable rates decrease faster since the Gaussian beam with a larger waist radius suffer more from the beam divergence induced by the increased transmission distance.

\begin{figure}[htbp]
\centering
\includegraphics[width=8.8cm]{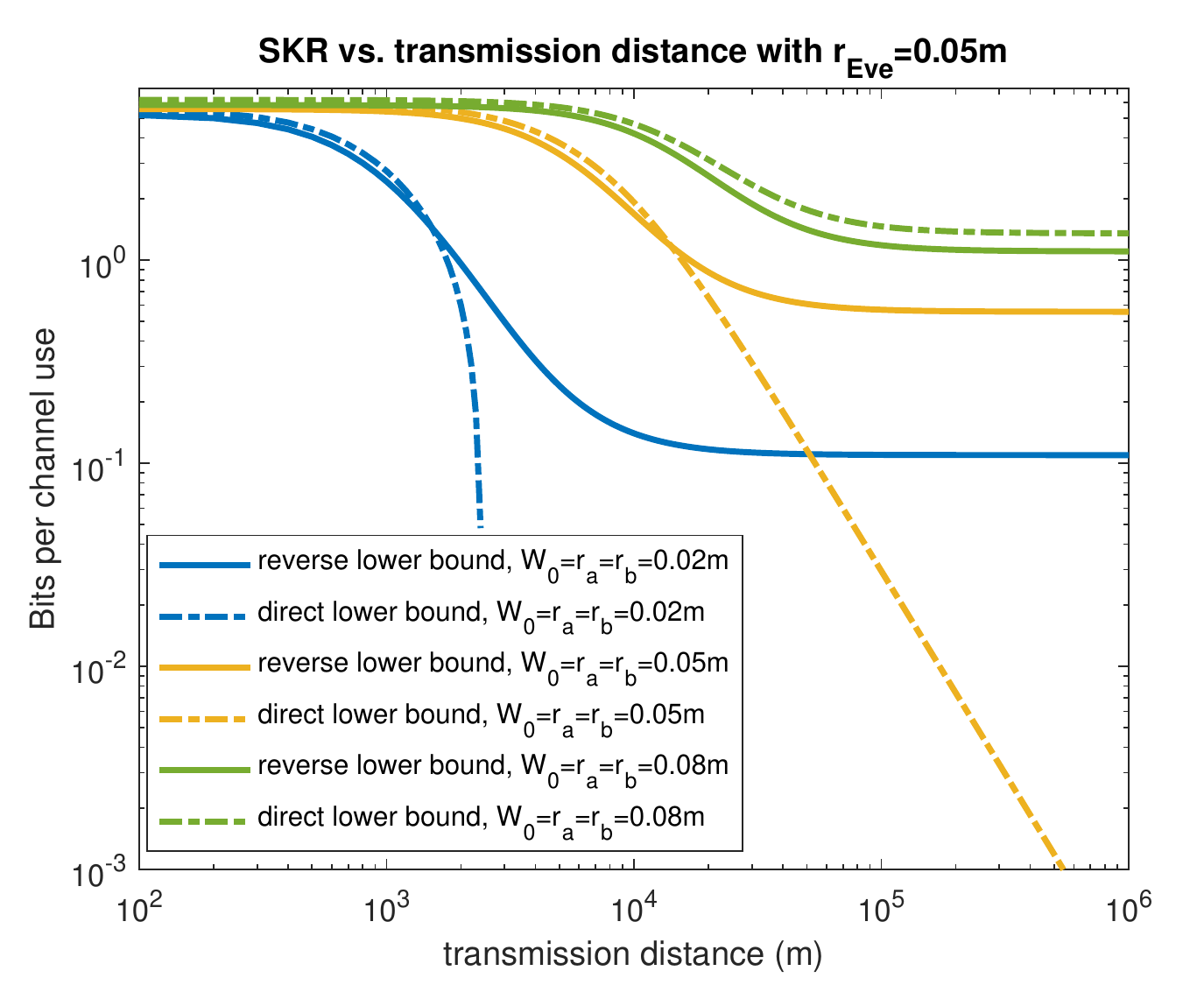}
\caption{Direct and reverse reconciliation SKR lower bounds vs. transmission distance $L$ with optimized input power (infinity as reconciliation efficiency $\beta$ is set to 1). 
 Radii of transmitted Gaussian beam waist and Bob's aperture are specified in the legend.  Transmission center wavelength $\lambda$ is set to 1550nm.  Eve's aperture radius is set to $r_e=5$cm. 
 \label{ConstantRatewithrarb}}
\end{figure}

From Fig.~\ref{ConstantRatewithrb} we know that increasing Bob's aperture size can always help increase the SKR achievable rate. And from Fig.~\ref{ConstantRatewithra} we know that increasing Alice's aperture size can help increase the SKR achievable rate when the transmission distance is large. In Fig.~\ref{ConstantRatewithrarb}, with $r_a$ and $r_b$ both increased at the same time, we plot both direct and reverse reconciliation achievable rates as functions of transmission distance. with $r_b=r_a=W_0$. We can see that when Bob's aperture size also increases, the achievable rate can be further increased compared to Fig.~\ref{ConstantRatewithra}. 

\begin{figure}[htbp]
\centering
\includegraphics[width=8.8cm]{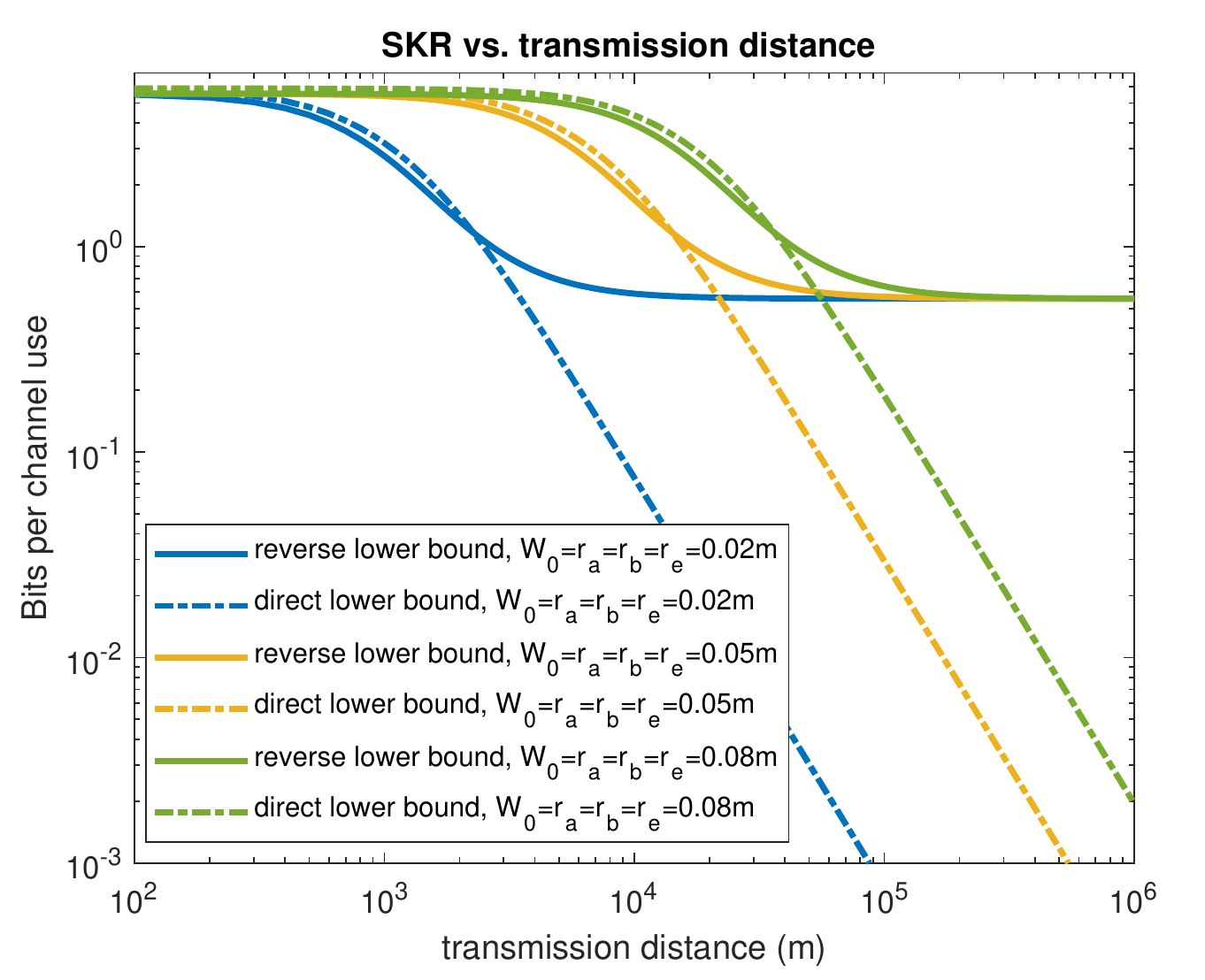}
\caption{Direct and reverse reconciliation SKR lower bounds vs. transmission distance $L$ with optimized input power (infinity as reconciliation efficiency $\beta$ is set to 1). 
 Radii of transmitted Gaussian beam waist and Bob's aperture are specified in the legend.  Transmission center wavelength $\lambda$ is set to 1550nm.   
 \label{ConstantRatewithrarbre}}
\end{figure}

In Fig.~\ref{ConstantRatewithrarbre} we further investigate the case when Alice's, Bob's and Eve's aperture sizes all increase at the same time. This is based on the assumption that any commercialized aperture on the market should be available to both the communication parties and the eavesdropper at the same time. We use this assumption to test how the same technology advance would affect the communication parties and the eavesdropper. We can see that the SKR lower bound curves basically move horizontally along the x-axis, meaning that the SKR achievable rate improves against long transmission distance. However, since $r_b/r_e$ remains the same, the converging rate as transmission distance goes to infinity doesn't change.

\begin{figure}[htbp]
\centering
\centering
\includegraphics[width=9cm]{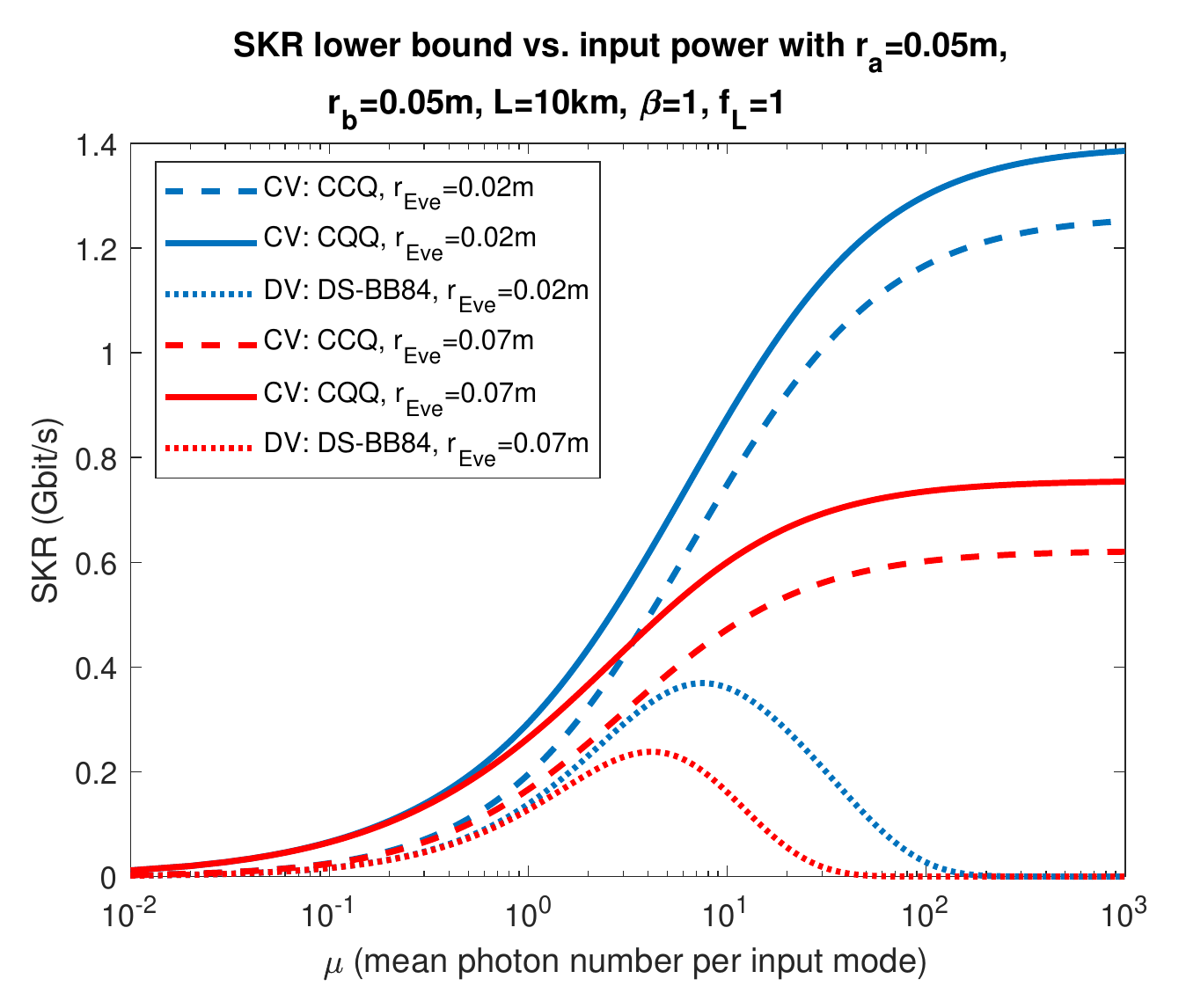}
\caption{SKR lower bound for Gaussian modulated CV-QKD and DS-BB84 vs. input power with perfect information reconciliation ($\beta=1, f_L=1$). CQQ rates are also included for comparison.
The transmission distance $L$ is 100km. Radius of Eve aperture ($r_{e}$) are specified in the legend. Transmitted Gaussian beam waist radius is set to $W_0=5$cm. Bob and Alice aperture radius are set to $r_b=r_a=5$cm. R=1Gbit/s. 
Transmission center wavelength $\lambda$ is set to 1550nm. \label{ComparisonVmuLA_cropped}}
\end{figure}

Below we present comparison results between Gaussian-modulated CV-QKD  (with coherent states, heterodyne detection and reverse reconciliation) and DS-BB84 protocols under the limited aperture size scenario. A weak coherent-state source is assumed to be Alice's transmission source and signal-state pulses are transmitted from Alice to Bob containing $\mu$ mean photons per pulse at a rate of $R$ states/s. For CV-QKD we use the CCQ (classical-classical-quantum) rate (solid curve) as in Eq.~(69) from~\cite{pan2019secret} and for DS-BB84 (dotted curves) we use Eq.~(95) with reconciliation efficiency $f_L$ from~\cite{pan2019secret}. We also include lower bounds obtained with Eqs.~(\ref{LBRmu}) as CQQ (classical-classical-quantum) rate (dashed curves). Here CQQ indicates that one of the communication parties, Alice or Bob, has performed a measurement on her/his quantum system, while the other and Eve are yet to measure their respective quantum systems. On the other hand, CCQ indicates that both Alice and Bob have performed measurements~\cite{pan2019secret}. 

In Fig.~\ref{ComparisonVmuLA_cropped} we plot the comparison between CV and DV protocols   with perfect reconciliation ($\beta=1$, $f_L=1$).  Here different radii of Eve's aperture are included for comparison. We can see that the CCQ rate is lower than the CQQ rate for the CV protocol. It is also shown here in Fig.~\ref{ComparisonVmuLA_cropped} that the DS-BB84 rate is less than the CV rates under perfect reconciliation. 

\subsection{Imperfect Reconciliation Scheme}\label{ImperfectReconciliationScheme}
In this subsection we optimize the input power for the case with imperfect reconciliation ($\beta<1$). Below in Fig.~\ref{ImperfectRecon202006081232} (a) we plot the SKR lower bounds as functions of transmission distance with center wavelength $\lambda=1550$nm, $r_b=r_a=W_0=5$cm, $r_e=10$cm where we optimize input power for any given transmission distance as is shown in Fig.~\ref{ImperfectRecon202006081232} (b). We can see that Fig.~\ref{ImperfectRecon202006081232} (a) shows "constant rates" when $L$ is large similar to Fig.~\ref{ConstantRate} where reconciliation is perfect ($\beta=1$). In Fig.~\ref{ImperfectRecon202006081232} (b) we can see that the optimal input power changes with the transmission distance in a similar manner regardless of specific reconciliation efficiency. It is worth noticing that there is a transmission distance where the optimal input power is the lowest in Fig.~\ref{ImperfectRecon202006081232} (b), which is around the same transmission distance where the lower bounds start to converge to constant values in Fig.~\ref{ImperfectRecon202006081232} (a). In Fig.~\ref{ImperfectRecon202006081232} (b) we can also see that the optimal input power corresponding to given transmission distance increases with increasing reconciliation efficiency, which leads to higher SKR lower bounds in Fig.~\ref{ImperfectRecon202006081232} (a). 
\begin{figure}[H] 
\centering{\includegraphics[width=0.5\textwidth]{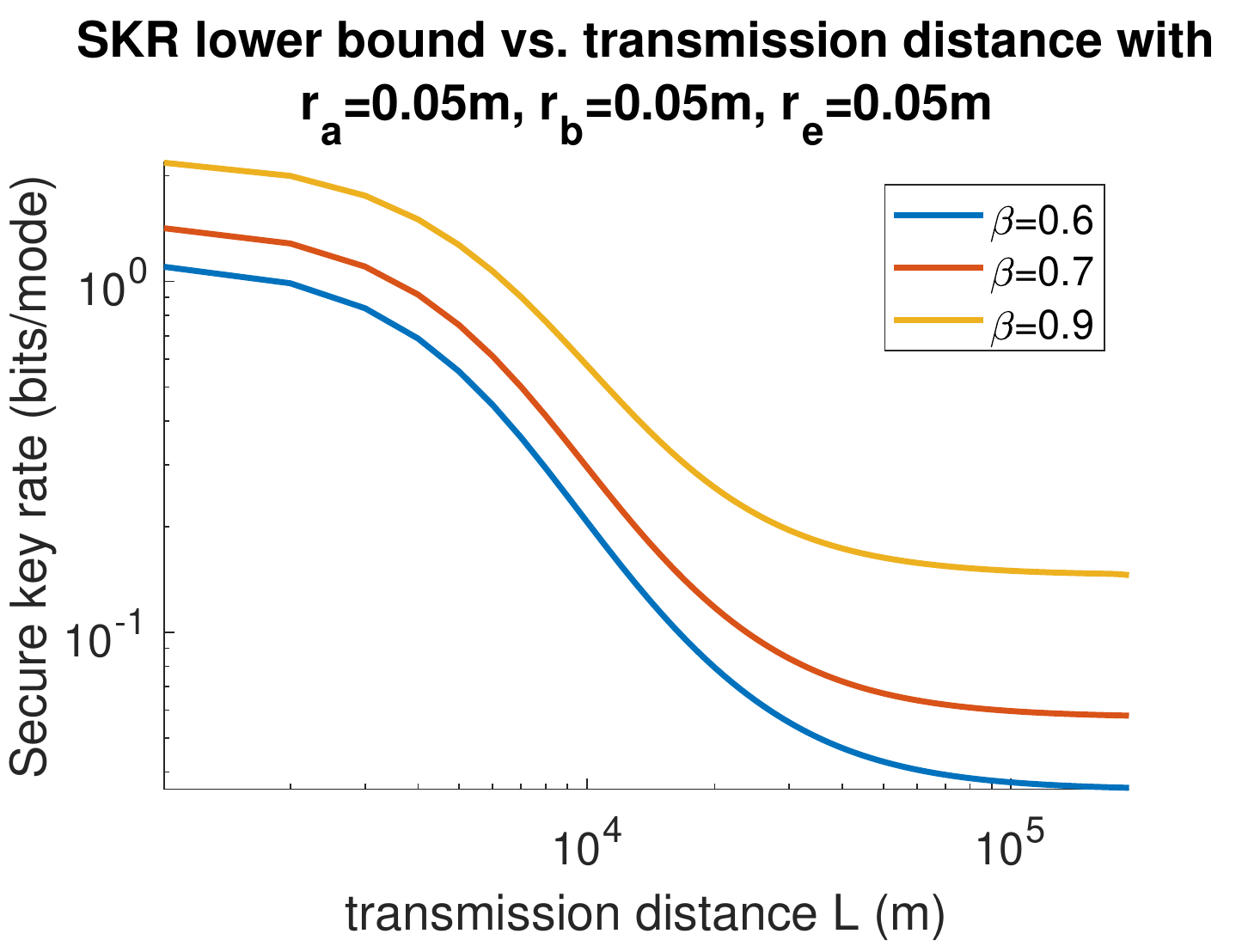}\\
\includegraphics[height=0.8pc]{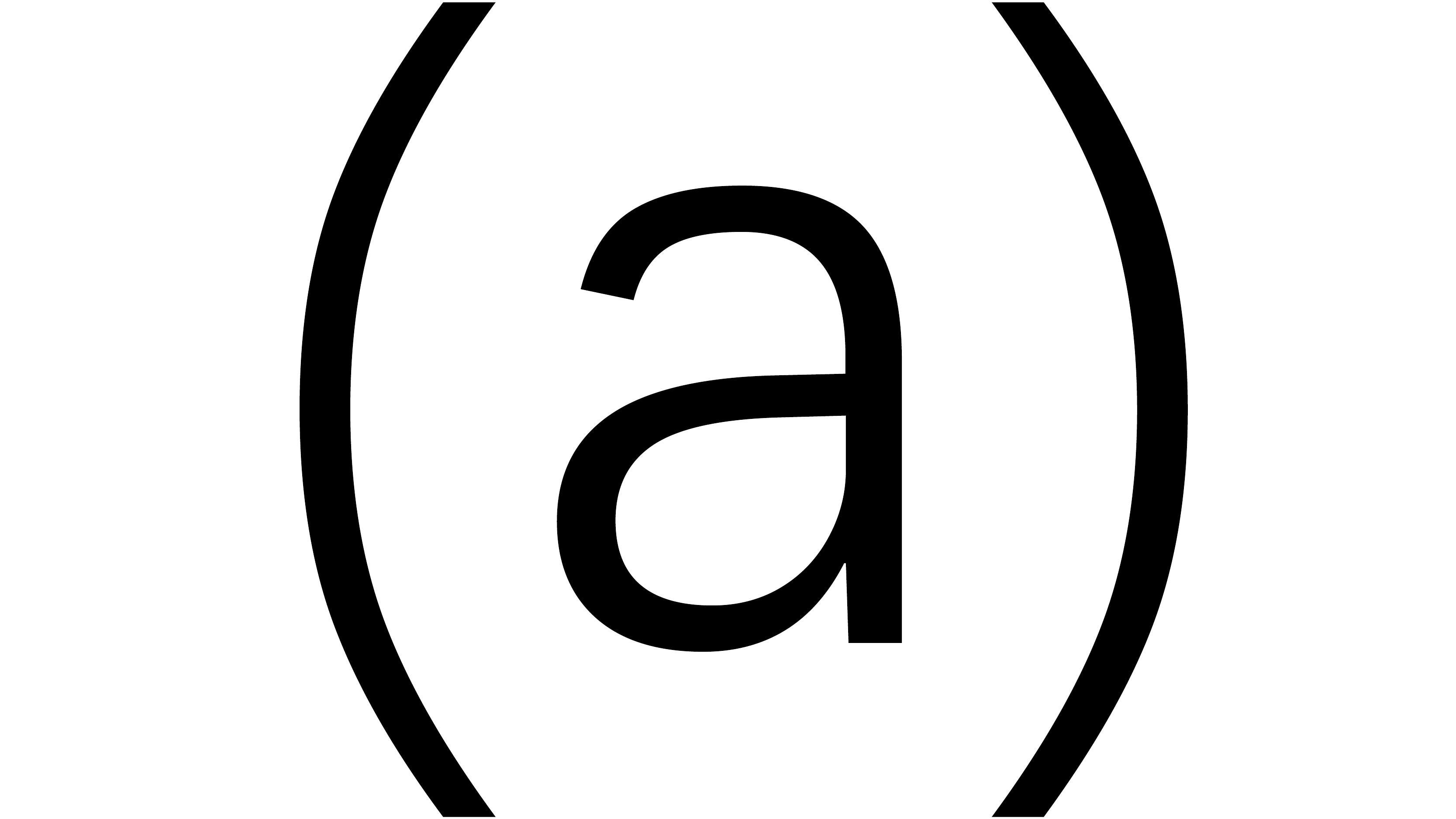}}\\
{\includegraphics[width=0.5\textwidth]{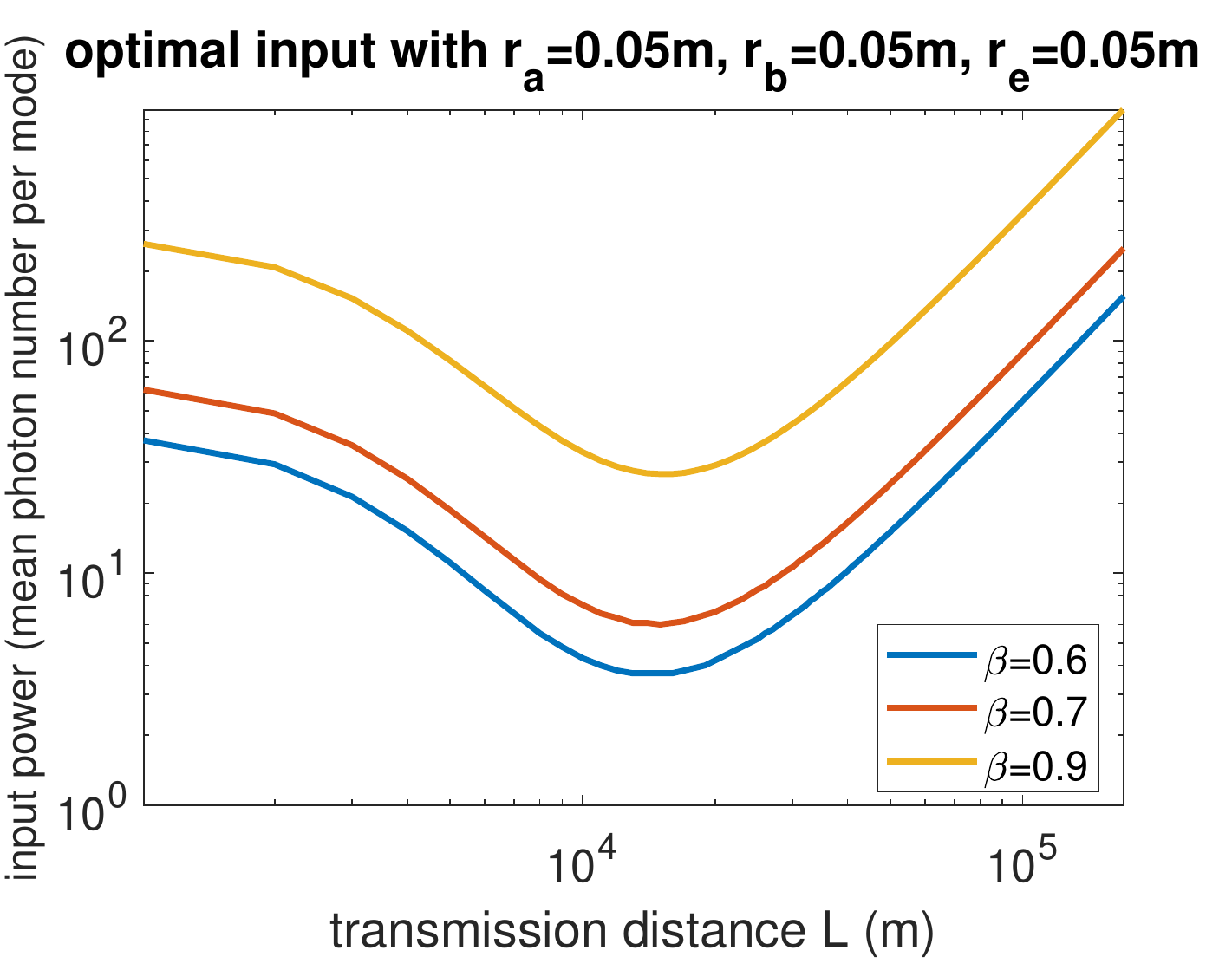}\\
\includegraphics[height=0.8pc]{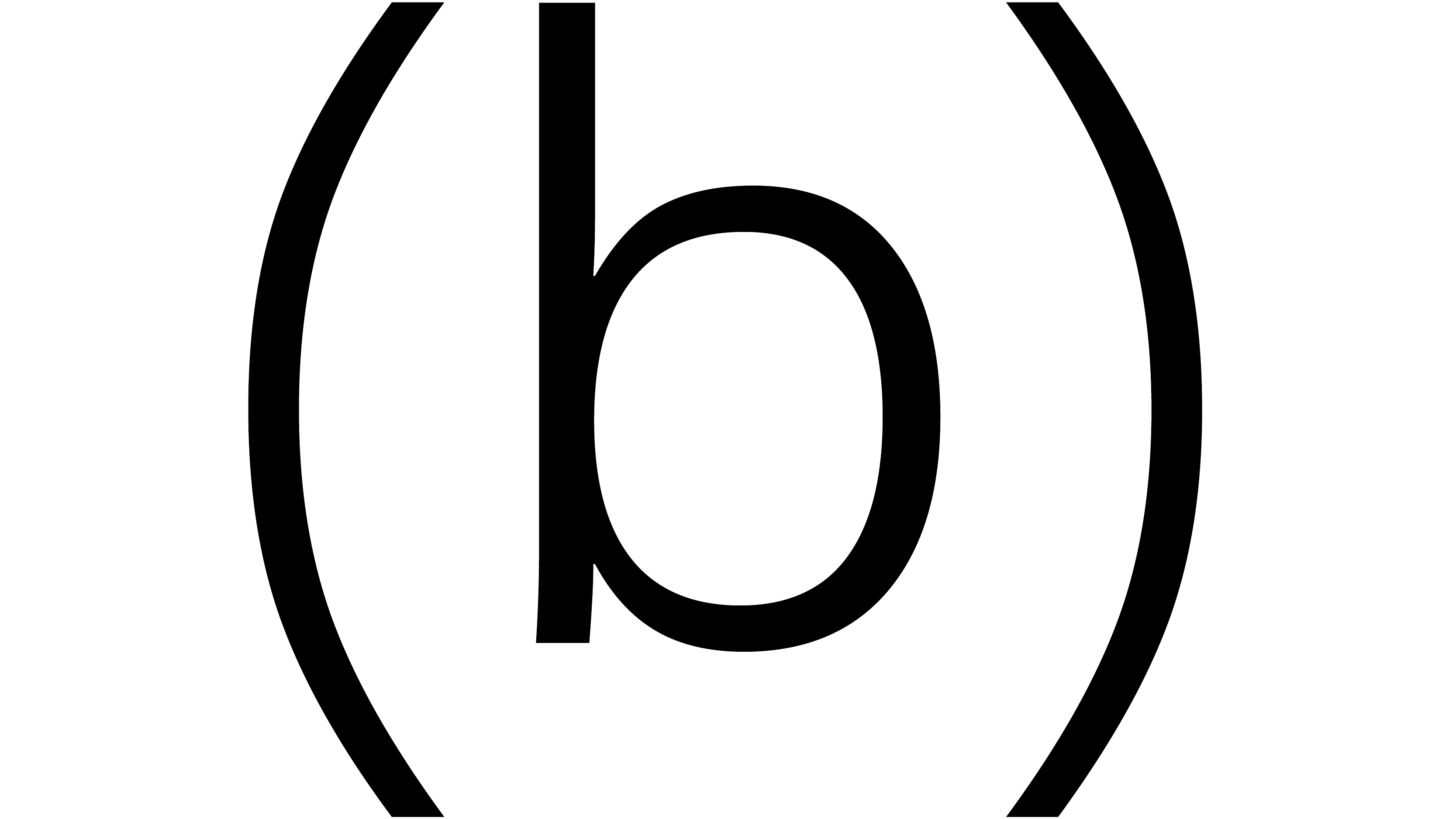}}
\caption{(a) SKR lower bound vs. transmission distance $L$ with optimized input power. Reconciliation efficiency $\beta$ are varied as given.
  Transmission center wavelength $\lambda$ is set to 1550nm. Transmitted Gaussian beam waist radius is set to $W_0=5$cm. Bob, Alice and Eve aperture radius are set to $r_b=r_a=r_e=5$cm. (b) Corresponding optimal input power for Fig.~\ref{ImperfectRecon202006081232} (a).\label{ImperfectRecon202006081232}}
\end{figure}

\begin{figure}[h]
\centering
\centering
\includegraphics[width=7.2cm]{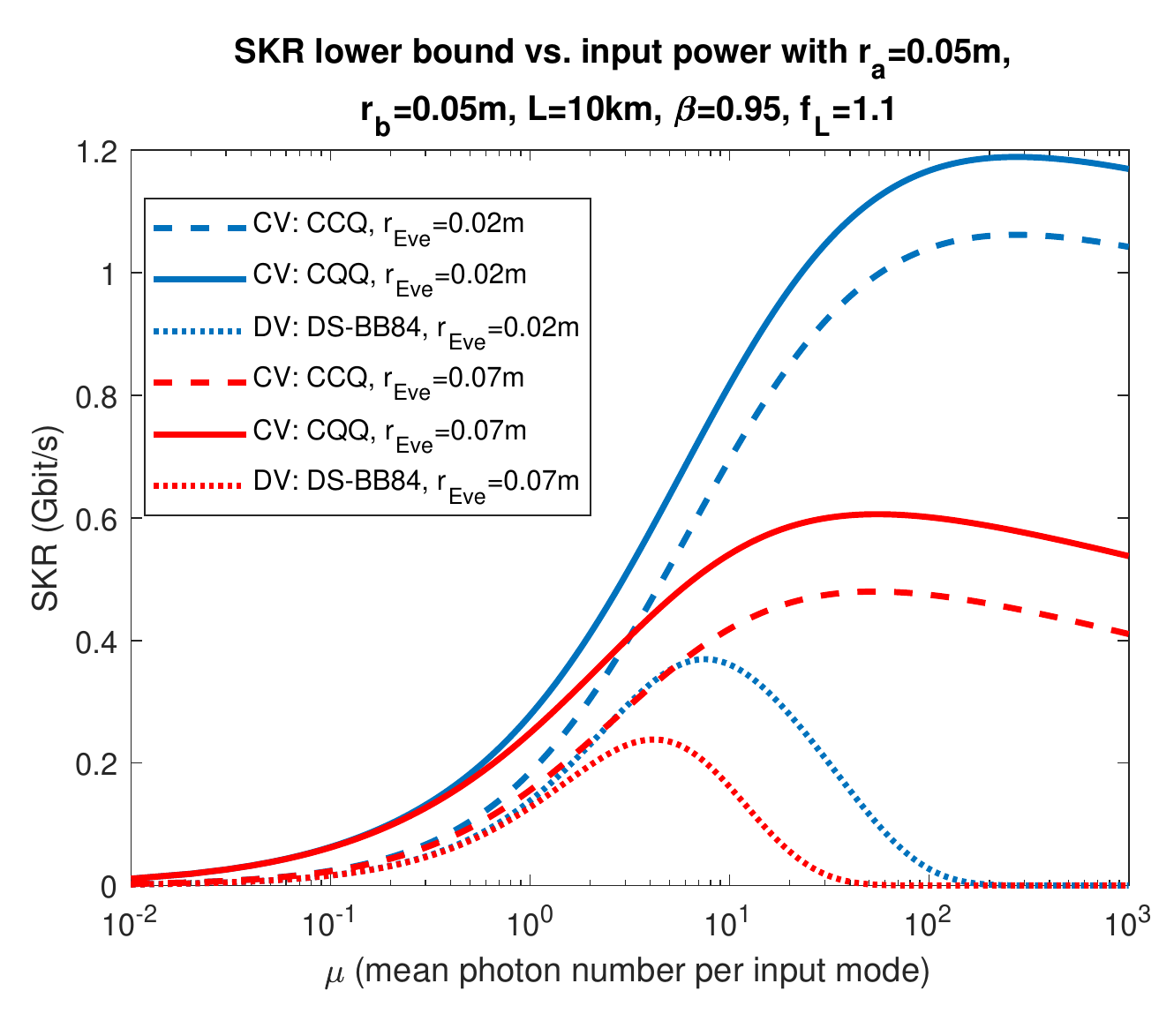}
\caption{SKR lower bound for Gaussian modulated CV-QKD and DS-BB84 versus input power with imperfect information reconciliation ($\beta=0.95, f_L=1.1$). CQQ rate are also included for comparison.
The transmission distance $L$ is 100km. Radius of Eve aperture ($r_{e}$) are specified in the legend. Transmitted Gaussian beam waist radius is set to $W_0=5$cm. Bob and Alice aperture radius are set to $r_b=r_a=5$cm. R=1Gbit/s. 
Transmission center wavelength $\lambda$ is set to 1550nm. \label{ComparisonVmuLAbeta9_cropped}}
\end{figure}

We further plot the comparison between CV and DV protocols in  Fig.~\ref{ComparisonVmuLAbeta9_cropped} with imperfect ($\beta=0.9$, $f_L=1.1$) reconciliation. 
Although in Fig.~\ref{ComparisonVmuLA_cropped} the DS-BB84 rate is less than the CV rates under perfect reconciliation, here in Fig.~\ref{ComparisonVmuLAbeta9_cropped}  DS-BB84 rate can exceed CV rate with some input power. However it still cannot exceed CV rate with optimized input power.

\begin{figure}[H] 
\centering{\includegraphics[width=0.45\textwidth]{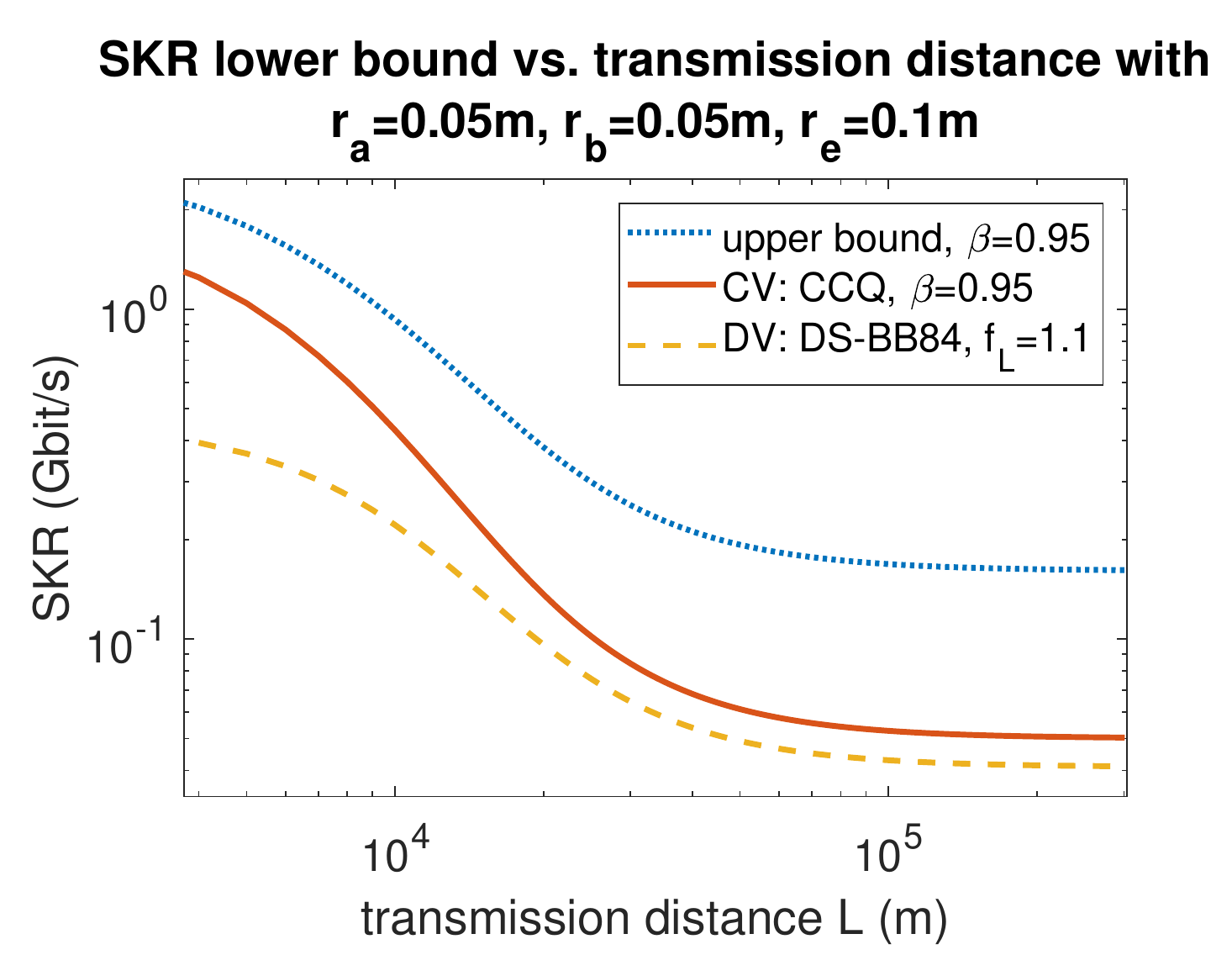}\\
\includegraphics[height=0.8pc]{a.pdf}}\\
{\includegraphics[width=0.45\textwidth]{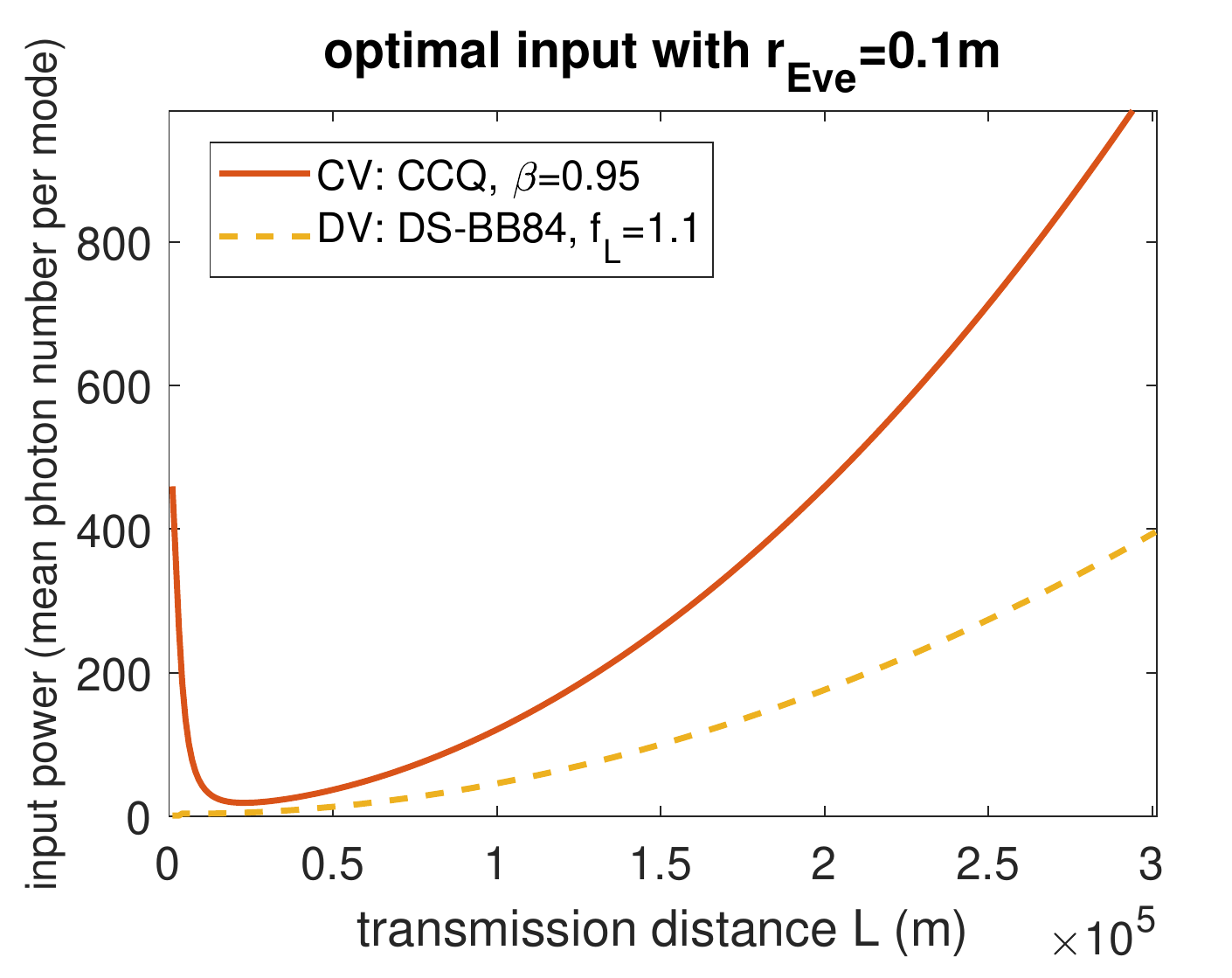}\\
\includegraphics[height=0.8pc]{b.pdf}}
\caption{(a) SKR lower bound for Gaussian modulated CV-QKD and DS-BB84 versus transmission distance $L$ with optimized input power. Upper bound is also included for comparison. Reconciliation efficiency are set as $\beta=0.95, f_L=1.1$.
  Transmission center wavelength $\lambda$ is set to 1550nm. Transmitted Gaussian beam waist radius is set to $W_0=5$cm. Bob, Alice and Eve aperture radius are set to $r_b=r_a=r_e=5$cm. R=1Gbit/s. (b) Corresponding optimal input power for Gaussian modulated CV-QKD and DS-BB84 in Fig.~\ref{CompOptimalLA} (a).\label{CompOptimalLA}}
\end{figure}

Next we focus on the comparison between CV-CCQ rate and DS-BB84 rate lower bounds where we also include the upper bound from Eq.~(\ref{UpBREEPL}). In Fig.~\ref{CompOptimalLA} (a) we plot the SKR lower bounds as functions of transmission distance with center wavelength $\lambda=1550$nm, $r_b=r_a=W_0=5$cm, $r_e=10$cm. The corresponding optimal input power is shown in Fig.~\ref{CompOptimalLA} (b) for both CV-CCQ rate and DS-BB84 rate. Here we can see that Fig.~\ref{CompOptimalLA} (a) shows a constant rate convergence similar to Fig.~\ref{ConstantRate} where CV-CCQ rate is always higher than DS-BB84. 

\begin{figure}[H] 
\centering{\includegraphics[width=0.45\textwidth]{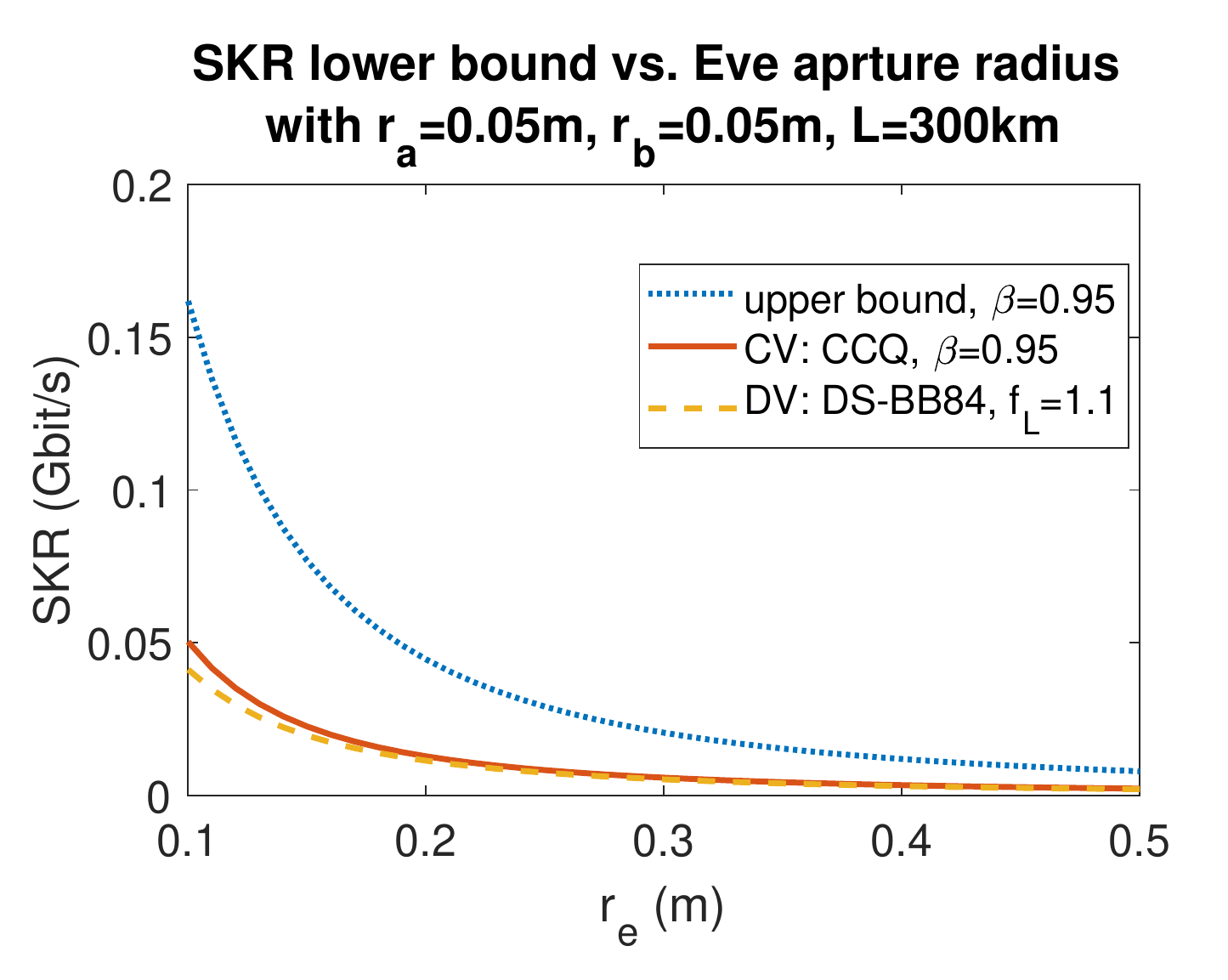}\\
\includegraphics[height=0.8pc]{a.pdf}}\\
{\includegraphics[width=0.45\textwidth]{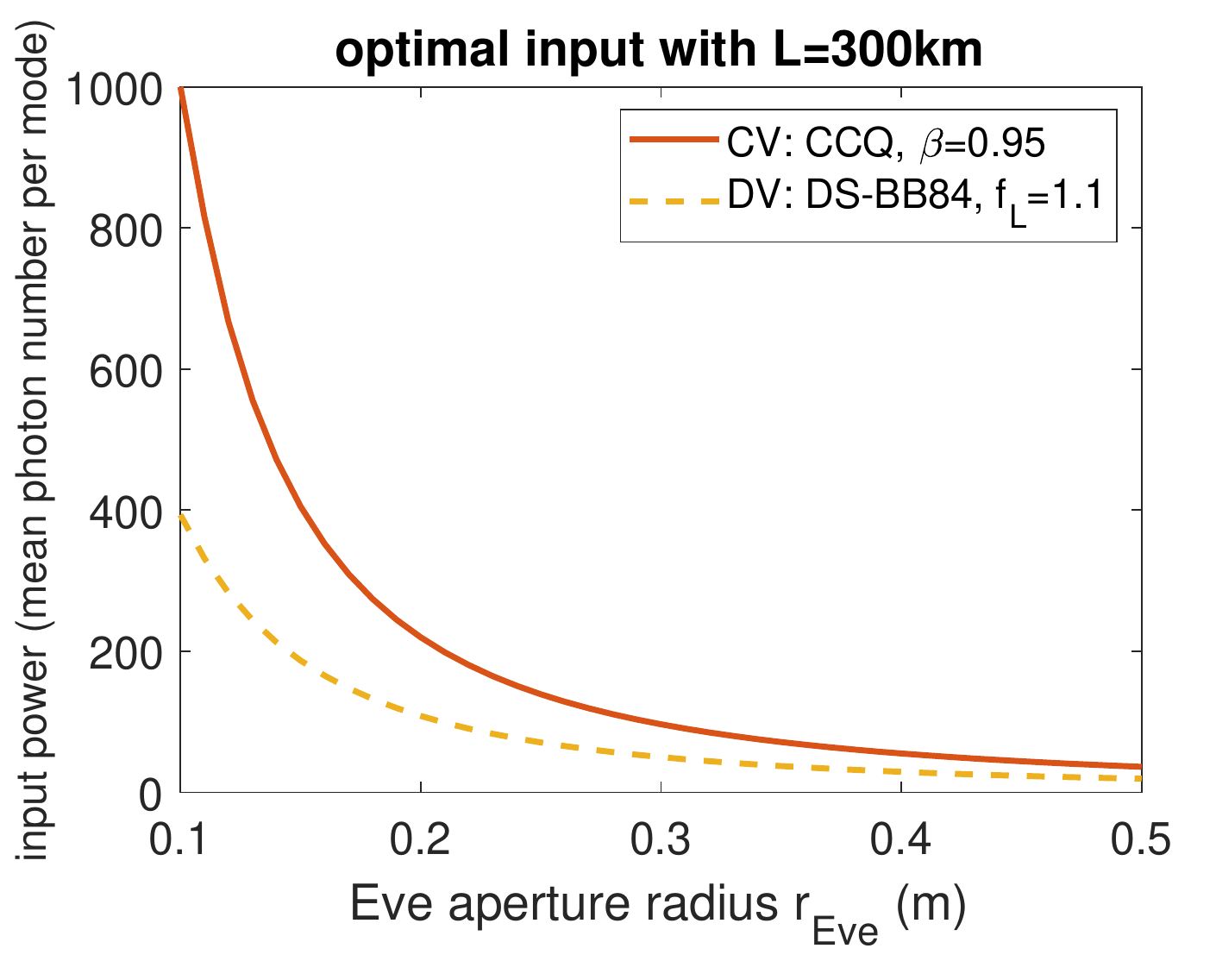}\\
\includegraphics[height=0.8pc]{b.pdf}}
\caption{(a) SKR lower bound for Gaussian modulated CV-QKD and DS-BB84 versus Eve aperture radius $r_e$ with optimized input power. Upper bound is also included for comparison. Reconciliation efficiency are set as $\beta=0.95, f_L=1.1$. Transmission distance is set as $L=300$km.
  Transmission center wavelength $\lambda$ is set to 1550nm. Transmitted Gaussian beam waist radius is set to $W_0=5$cm. Bob and Alice  aperture radius are set to $r_b=r_a=5$cm. R=1Gbit/s. (b) Corresponding optimal input power for Gaussian modulated CV-QKD and DS-BB84 in Fig.~\ref{CompOptimalvreLA} (a).\label{CompOptimalvreLA}}
\end{figure}

In Fig.~\ref{CompOptimalvreLA} (a) we optimize the input power and plot the SKR lower bounds as functions of Eve's aperture radius $r_e$ with center wavelength $\lambda=1550$nm, $r_b=r_a=W_0=5$cm, transmission distance $L=300$km. The corresponding optimal input power are shown to decrease with increasing $r_e$ in Fig.~\ref{CompOptimalvreLA} (b) for both CV-CCQ rate and DS-BB84 rate. Here we can see that both rates decrease with increasing $r_e$ with DS-BB84 rate approximating CV-CCQ rate from below.

\section{Summary}


\noindent In this paper, we have analyzed SKR lower and upper bounds with respect to relevant channel parameters for a specific realistic scenario in free-space optics satellite-to-satellite secret key distillation where Eve only has a limited-sized aperture in the same plane of Bob. We have shown the input power dependency in this scenario with perfect and imperfect reconciliation as well as how Eve's aperture size impacts on the optimal input power. For perfect reconciliation schemes, we  found out that when Bob's aperture is comparable to Eve's we can get a distance independent SKR at a sufficiently large transmission distance and we have 
derived analytical expressions for both the lower bound and upper bound with respect to direct and reverse reconciliation. We also showed similar phenomena with imperfect reconciliation and presented comparison between Gaussian-modulated CV-QKD and DS-BB84 protocols under these conditions. 
%

\section*{Acknowledgement}
This paper was supported in part by L3Harris, MURI ONR, NSF, and GD. The authors thankfully acknowledge helpful discussions with Saikat Guha, Kaushik Seshadreesan and John Gariano from the University of Arizona, Jeffrey Shapiro from Massachusetts Institute of Technology and William Clark, Mark R. Adcock from General Dynamics.

\bibliographystyle{IEEEtran}
\bibliography{QSDC}






\end{document}